\documentclass[showpacs,preprint,nofootinbib,superscriptaddress,citeautoscript,eqsecnum,12pt]{article}
\usepackage{latexsym}
\usepackage{amsmath}
\usepackage{amsfonts}
\usepackage{amssymb}
\usepackage{amscd}
\usepackage{bbm}
\usepackage{fancybox}
\usepackage{cite}
\usepackage{amsmath,amsfonts,amsbsy}
\usepackage{pstricks,pst-node}
\usepackage{textcomp}
\usepackage{mathrsfs}
\usepackage{multirow}
\usepackage{array}
\usepackage{tensor}

\pdfoutput=1
\usepackage{graphicx}
\usepackage{color}
\usepackage{epsfig}
\usepackage{psfrag}
\usepackage{comment}
\usepackage{epstopdf}
\usepackage{hyperref}

\usepackage[utf8x]{inputenc}
\usepackage{relsize}
\usepackage{dcolumn}

\usepackage{float}

\psset{unit=1.3cm,linewidth=.5pt,radius=.2}  % controls the size of diagrams

\usepackage{multirow}                     % tables cells spanning multiple rows
\usepackage{float}                          % to force floaters to stay Here
\usepackage{lscape}                         % landscape laid out pages
\usepackage{subfigure}          % for subfigures
\usepackage{bm}

\newcommand{\cA}{{\cal A}}
\newcommand{\cB}{{\cal B}}

\newcommand{\cH}{{\cal H}}

\newcommand{\cL}{{\cal L}}
\newcommand{\cM}{{\cal M}}
\newcommand{\cN}{{\cal N}}
\newcommand{\cO}{{\cal O}}
\newcommand{\cP}{{\cal P}}

\newcommand{\cR}{{\cal R}}

\newcommand{\cW}{{\cal W}}

\newcommand{\bb}{\bar\beta}

\newcommand{\beq}{\begin{equation}}
\newcommand{\eeq}{\end{equation}}
\newcommand{\bi}{\begin{itemize}}
\newcommand{\ei}{\end{itemize}}
\newcommand{\bt}{\begin{tabular}}
\newcommand{\et}{\end{tabular}}
\newcommand{\bc}{\begin{center}}
\newcommand{\ec}{\end{center}}

\def\one{{\hbox{ 1\kern-.8mm l}}}
\newcommand{\Dslash}{\not{\hbox{\kern-4pt $D$}}}
\newcommand{\pdslash}{\not{\hbox{\kern-2pt $\partial$}}}
\newcommand{\nn}{\nonumber}

\DeclareMathOperator{\ad}{ad}
\DeclareMathOperator{\Ad}{Ad}

\newcommand{\be}{\begin{equation}}
\newcommand{\ee}{\end{equation}}
\newcommand{\bea}{\begin{eqnarray}}
\newcommand{\eea}{\end{eqnarray}}
\newcommand{\ba}{\begin{array}}
\newcommand{\ea}{\end{array}}

\def\bbox{{\,\lower0.9pt\vbox{\hrule \hbox{\vrule height 0.2 cm
\hskip 0.2 cm \vrule height 0.2 cm}\hrule}\,}}
\newcommand{\dsl}{\pa \kern-0.5em /}

\newcolumntype{C}[1]{>{\centering}m{#1}}
\newcommand{\bigzero}{\mbox{\normalfont\Large\bfseries 0}}
\newcommand{\rvline}{\hspace*{-\arraycolsep}\vline\hspace*{-\arraycolsep}}

\newcommand{\Ab}{\bar A}

\newcommand{\tr}{\mathrm{tr}}
\newcommand{\Str}{\mathrm{Str}}
\newcommand{\vp}{\varphi}

\newcommand{\ve}{\varepsilon}
\newcommand{\extd}{\mathrm{d}}
\newcommand{\umat}{\mathfrak{m}}

\font\mybb=msbm10 at 12pt
\def\bb#1{\hbox{\mybb#1}}

\def\bR {\bb{R}}

\definecolor{rr}{RGB}{204, 0, 0}

\begin{document}

\begin{titlepage}%1
\begin{center}

%\hfill UG-14-XX

\vskip 1.5cm

%\noindent
%{\Large \bf AdS$_3$ (super)gravity as the geometric action on the coadjoint orbit of (super-)Virasoro}

%\noindent
%{\Large \bf AdS$_3$ supergravity and Superreparameterizations}

\noindent
{\Large \bf  Asymptotic dynamics of three dimensional supergravity and higher spin gravity revisited}

\vskip 1cm

{\bf Wout Merbis$^{a}$}, {\bf Turmoli Neogi$^{b}$} and {\bf Arash Ranjbar$^{c}$}

\vskip 25pt

{\em 
$^a$ Dutch Institute for Emergent Phenomena \& Institute for Theoretical Physics, University of Amsterdam, Science Park 904, 1098 XH Amsterdam, The Netherlands. \vskip 5pt
$^b$ Centro de Estudios Cient\'{\i}ficos (CECs), Av. Arturo Prat 514, Valdivia, Chile. \vskip 5pt
$^c$ Division of Theoretical Physics, Rudjer Bo\v skovi\'c Institute \\ Bijeni\v cka 54, 10000 Zagreb, Croatia \vskip 5pt}
{email: {\tt  w.merbis@uva.nl, \\ ~~~~turmolineogi@yahoo.com, aranjbar@irb.hr}} \\
\vskip 10pt

\end{center}

\vskip 1.5cm

\begin{center} {\bf ABSTRACT}\\[3ex]
\end{center}

\noindent
We reconsider the Hamiltonian reduction of the action for three dimensional AdS supergravity and $W_3$ higher spin AdS gravity in the Chern-Simons formulation under asymptotically anti-de Sitter boundary conditions. We show that the reduction gives two copies of chiral bosons on the boundary. In particular, we take into account the holonomy of the Chern-Simons connection which manifests itself as zero mode of the momentum of the boundary chiral boson. We provide an equivalent formulation of the boundary action which we claim to be the geometric action on symplectic leaves of a (super-)Virasoro or a higher spin $W_N$ Poisson manifold in the case of supergravity or higher spin gravity respectively, where the intersection of leaves (given in terms of leaves representatives) can be identified as the bulk holonomy. This concludes the extension to non-linear algebras where the notion of coadjoint representation is not well-defined. The boundary Hamiltonian depends on a choice of boundary conditions and is equivalent to the Schwarzian action for corresponding Brown-Henneaux boundary conditions. We make this connection explicit in the extended supersymmetric case. Moreover, we discuss the geometric action in the case of $W_3$ AdS$_3$ gravity in both $\mathfrak{sl}(3)$ highest weight representations based on principal and diagonal $\mathfrak{sl}(2)$ embeddings.

\end{titlepage}

\tableofcontents

%%%%%%%%%%%%%%%%%%%%%%%%%%%%%%%
%%%%%%%%%%%%%%%%%%%%%%%%%%%%%%%
\section{Introduction}

The theory of gravity in three dimensions has the remarkable feature that it does not carry any dynamical degrees of freedom in the bulk. It is by now well known that general relativity in three dimensions is a topological gauge theory which can be described as a Chern-Simons theory for the appropriate gauge (super)group \cite{Achucarro:1987vz,Witten:1988hc}. The gauge group reflects the isometries of the maximally symmetric vacuum of the theory, which depends on the presence and sign of the cosmological constant $\Lambda$. In the present work, we will focus on negative cosmological constant $\Lambda = - 1/\ell^2$, mainly due to the fact that the presence of BTZ black hole solutions of \cite{BTZ} require a non-vanishing negative $\Lambda$. In three dimensional gravity, all dynamical degrees of freedom live on the asymptotic boundaries. This interesting property of the theory has spiked the interest in the theory as a candidate for a consistent theory of quantum gravity in three dimensions \cite{Witten:1988hc,Strominger:1997eq}. However, a direct computation of the partition function of AdS$_3$ gravity remained cumbersome or inconclusive \cite{Witten:2007kt,Maloney:2007ud}. 

Recently in \cite{Henneaux:2019sjx}, it was shown that the complete Poisson structure at the boundary is not given solely in terms of Kac-Moody currents, but also includes the zero modes of bulk holonomies, which can be understood as Wilson lines stretched between the boundaries. Indeed, these Wilson lines are classically traversable wormholes \cite{Maldacena:2001kr}. In the quantum theory and in the presence of two boundaries, this is understood as a thermofield double description of two entangled conformal field theory \cite{Guica:2015zpf,Harlow:2018tqv}. This means that the quantum description of the theory is based on a complete set of observables which are formed by the currents and the Wilson lines/loops. The Hilbert space does not factorize into two copies of boundary states, but the boundary theories are coupled through global zero modes.
In \cite{Henneaux:2019sjx}, a complete analysis was provided of such Wilson lines in pure Chern-Simons gravity on asymptotically AdS geometries with two boundaries, as the simplest example. It was shown that, in the presence of non-trivial holonomies in the bulk, the action obtained from a Hamiltonian reduction of the 3D Einstein-Hilbert action is exactly the geometric action on the Virasoro coadjoint orbit, with the holonomy being the orbit representative. 

These results can be promoted to three dimensional supergravity models, because these are also Chern-Simons theories with boundary conditions of the Drinfeld-Sokolov type, implementing a Hamiltonian reduction at the boundary \cite{Henneaux:1999ib}.  The resulting asymptotic symmetry algebras are the $\cN$-extended superconformal algebras of \cite{Knizhnik:1986wc,Bershadsky:1986ms,Fradkin:1992bz,Fradkin:1992km}, which are linear for $\cN \leq 2$. A similar discussion in two dimensions has been considered for the BF formulation of dilaton supergravity \cite{Cardenas:2018krd}.  

In the case of an annulus topology, we include the holonomies along the lines of \cite{Henneaux:2019sjx,Henneaux:1999ib} by treating separately the two chiralities. For each chirality, it leads to a supersymmetric chiral action at each boundary coupled by radial Wilson lines.
One also finds that the system is physically described by two sets of generators of the superconformal algebras, one at each boundary.  These generators are constrained by the holonomy matching condition and provide, together with the global modes, a complete description of the system. The dynamics reduces to the dynamics of these generators and of the global modes, and can therefore be expressed in terms of geometric actions. 

In the geometric construction of the action based on Kirillov-Kostant prescription \cite{Kirillov1976,Kostant1970,Kirillov2004}, the existence of a coadjoint representation is an utmost necessity. However, this notion is not sufficient and will be challenged when the boundary algebras are non-linear. This is the case, for example, when the asymptotic charges of the theory form a non-linear algebra as in supergravity with $\mathcal{N}>2$ or higher spin gauge theories with non-linear $\cW$-algebras \cite{Henneaux:2010xg,Campoleoni:2010zq}. Then, the geometric actions can not be formulated in terms of orbits of the coadjoint representation, since the phase space does not provide a linear representation, but rather in terms of the more general concept of symplectic leaves \cite{Lichnerowicz1977,Weinstein1983}. When the asymptotic symmetry algebra is linear, the boundary action can be cast in the form of the Schwarzian action \cite{Alekseev:1988ce,Alekseev:1990mp} which has been shown to be the geometric action on the coadjoint orbits of the asymptotic symmetry group. From the point of view of Hamiltonian reduction, the chiral boundary theory can be also shown to be a Schwarzian action where the holonomies appear as the constant orbit representatives.

The generators of the asymptotic symmetry algebra form a Poisson manifold, with a Poisson bracket that is degenerate if one focuses only on a single boundary algebra without including the global radial Wilson lines. The symplectic leaves of this Poisson manifold have a well-defined symplectic structure, which is the one that enters in the action. Moreover, it happens that the Hamiltonian reduction in the presence of non-trivial holonomy still indubiously provides the action on the boundary.  Therefore, one is able to perform a Hamiltonian reduction in supergravity or higher spin gravity in the presence of bulk holonomies in order to obtain the boundary theory. These boundary actions should be considered as a candidate for a geometric action on the symplectic leaves of the Poisson manifold. The holonomies are related to the leave representatives; in fact, the intersections of symplectic leaves are determined by the zero modes of asymptotic charges which in turn are given by the holonomies.

A great advantage of our approach is that holonomies appear in the action as dynamical fields. They are considered as time-dependent variables in the action. Their equations of motion set them on-shell to constant variables determined by the zero modes of the asymptotic charges. These constant holonomies match with the constant orbit representatives. This happens to be an important feature once one wants to quantize the theory, since in a genuine quantum theory of gravity one should consider summing over all possible solutions of the theory. The presence of holonomies in the action provides a control over the solution space at the level of action.

Recently, there has been a surge of interest in the computation of the path integral of two dimensional Jackiw-Teitelboim (JT) gravity. In two dimensions, the partition function of gravity turns out to have a beautiful mathematical description. It is the volume of the moduli space of hyperbolic manifolds with constant curvature and the boundary theory is an ensemble of one dimensional quantum mechanical models, which can be captured by random matrix theory. 
An important piece of information is that in two dimensions, the partition function is written as the exponential of a Schwarzian action functional \cite{Saad:2019lba}. This Schwarzian action is the boundary theory of 2D JT gravity and can be obtained as the IR limit of the SYK model \cite{Kitaev:2015,Sachdev:1992fk,Maldacena:2016hyu,Maldacena:2016upp}. In three dimensions, this piece of information happened to be crucial, as the Hamiltonian on the boundary of 3D Chern-Simons gravity is given by the sum of two Schwarzian actions with opposing chirality \cite{Mertens:2018fds,Alekseev:1990mp,Henneaux:2019sjx}.
There has been a plethora of works in 3D along this line of thoughts \cite{Cotler:2020ugk,Afkhami-Jeddi:2020ezh,Benjamin:2021wzr,Benini:2022hzx,Chandra:2022bqq,Mertens:2022ujr}.

In this work, we provide a systematic recipe to write down the Schwarzian action for 3D supergravity and higher spin gravity in the presence of non-trivial holonomy. While we pay special attention to the role of bulk holonomies, we focus on the case where the holonomy is in the hyperbolic conjugacy class. This is because we are mostly interested to exhibit the result for BTZ black hole solutions and these solutions are in the hyperbolic holonomy conjugacy class. Upon Hamiltonian reduction, the boundary Hamiltonian depends on a choice of boundary conditions and is equivalent to the Schwarzian action for Brown-Henneaux boundary conditions. This Schwarzian action may be considered as the starting point for computing the partition function of the corresponding theory. There is a benefit in using this action for computing the partition function. It already contains information on the Wilson lines, and therefore it may be more suited in addressing the factorization problem in 3D gravity. In this paper, we only focus on the construction of the boundary actions and details regarding computation of the partition function will be discussed elsewhere.

Our paper is organized as follows. In Section \ref{sec:AdS3-gen} we provide a short introduction to AdS$_3$ supergravity and $W_3$ higher spin theories in three dimensions. In each case, we discuss the gauge fixing and the corresponding Brown-Henneaux boundary conditions. 
In Section \ref{sec:reduction}, we perform the Hamiltonian reduction on the super Chern-Simons theory, or equivalently, AdS$_3$ supergravity. We first show this explicitly for three dimensional AdS supergravity with $\cN=1,2$ where one can use the Kirilov-Kostant construction to compare. The boundary action of Chern-Simons AdS$_3$ supergravity for an arbitrary number of supersymmetry $\cN$ is addressed in Section \ref{sec:N2superCS}. The honolomy part of the action remains undetermined for a general case and requires a case by case study. We will comment on that in Section \ref{sec:generalN}. 
Section \ref{sec:Higher-Spin} is aimed at the Hamiltonian reduction of $W_3$ AdS$_3$ higher spin theory. In this case, there are two distinct boundary conditions depending on the $\mathfrak{sl}(2)$ embedding of the $\mathfrak{sl}(3)$ algebra. We discuss both the principal and diagonal embedding and provide the boundary action in each case.

In Section \ref{sec:geometric-description}, we discuss the geometric action on the coadjoint orbit of the group of reparameterizations of the supercircle. We show that the result of Section \ref{sec:reduction} for $\cN=1,2$ is precisely the one of the geometric action. 
Appendix \ref{app:conv} provides some conventions regarding $Osp(2|2)$ representation, appendix \ref{app:sym_leaves} comprises a brief view of action on the symplectic leaves and appendix \ref{app:geomquan} is a review of geometric quantization for (super)conformal groups.

%%%%%%%%%%%%%%%%%%%%%%%%%%%%%%%
%%%%%%%%%%%%%%%%%%%%%%%%%%%%%%%
\section{Three dimensional AdS gravity}\label{sec:AdS3-gen}

\subsection{AdS$_3$ supergravity}\label{sec:AdS3}
Three dimensional AdS supergravity is described by a Chern-Simons theory for a Lie super-group $G$, where the even part of the group must contain $SO(2,2) \cong SL(2,\bR)\times SL(2,\bR)$. The Einstein Hilbert action can then be written as the difference of two Chern-Simons actions
\begin{equation}\label{Seh}
S_{\textsc{gr}}[\cA,\bar{\cA}] = S_{\textsc{cs}}[\cA] - S_{\textsc{cs}}[\bar{\cA}]\,,
\end{equation}
where
\begin{equation}\label{Scs}
S_{\textsc{cs}}[\cA] =  \frac{k}{4\pi} \int_{\cM} \Str\left( \cA \wedge d \cA + \frac23 \cA \wedge \cA \wedge \cA \right)\,,
\end{equation}
and $k = \frac{\ell}{4G_N}$. The gauge connections $\cA,\bar{\cA}$ take values in the superalgebra $\mathfrak{g}_0 \oplus \mathfrak{g}_1$ where  $\mathfrak{g}_0 = \mathfrak{sl}(2,\bR) \oplus \tilde{\mathfrak{g}}_0$. Here $\tilde{\mathfrak{g}}_0$ is the Lie algebra of the corresponding $R$-symmetry group. The $\mathfrak{sl}(2,\bR)$ generators are denoted by $L_{-1},L_0, L_{+1}$ and satisfy the commutation relations
\be\label{bracket-init}
[L_0,L_{\pm}] = \pm L_{\pm},\qquad [L_+,L_{-}]=2L_0.
\ee
When $G = OSp(\cN|2) \times OSp(\cN|2)$, the Lie algebra $\tilde{\mathfrak{g}}_0$ of $R$-symmetry is $\mathfrak{so}(\cN)$. Let's consider $T^a$ $(a=1,...,D)$ to be generators of $\mathfrak{so}(\cN)$ where $D=\cN(\cN-1)/2$ is the dimension of the algebra $\mathfrak{so}(\cN)$. They satisfy the following commutation relations
\begin{align}
[T^a,T^b] &= f^{ab}{}_c T^c,\\
[T^a, L_{\pm,0}] &= 0.
\end{align}
Fermionic generators transform in the fundamental representation of $SO(\cN)$. Therefore the odd part $\mathfrak{g}_1$ consists of $2\cN$ generators $Q_{\pm}^{\alpha}$ where $\alpha=1,...,d$. Note that $d=\cN$ is the dimension of fundamental representation of $SO(\cN)$ in which fermionic generators transform. The algebra $\mathfrak{osp}(\cN|2,\mathbb{R})$ is then defined as
\begin{subequations}
	\label{ospN}
\begin{align}
[L_0,Q_{\pm}^{\alpha}] &= \pm \frac{1}{2} Q_{\pm}^{\alpha},\\
[L_{\pm},Q_{\pm}^{\alpha}] &= 0,\\
[L_{\pm},Q_{\mp}^{\alpha}] &= + Q_{\pm}^{\alpha},\\
[T^a,Q_{\pm}^{\alpha}] &= - (\lambda^a){}^\alpha{}_\beta\, Q_{\pm}^{\beta},\\
\{Q_{\pm}^{\alpha},Q_{\pm}^{\beta}\} &= \pm \eta^{\alpha\beta} L_{\pm},\\
\{Q_{\pm}^{\alpha},Q_{\mp}^{\beta}\} &= - \eta^{\alpha\beta} L_{0} \pm \frac{d-1}{2 C_{\rho}} (\lambda^a)^{\alpha\beta} T^a \label{bracket-end}.
\end{align}   
\end{subequations}
Here $\lambda^a$ are the basis of the representation $\rho$ of $SO(\cN)$ in which fermions transform, $\eta^{\alpha\beta}$ is an $SO(\cN)$ invariant metric, $C_{\rho}$ is the second Casimir in the $\rho$ representation, i.e. one has
\be
\lambda^a \lambda^a = - C_{\rho} I, \qquad tr(\lambda^a \lambda^b) = - \frac{d}{D} C_{\rho} \delta^{ab}.
\ee
The graded super Jacobi identity for any $X,Y,Z \in \mathfrak{g}$ is written as
\begin{align}
(-1)^{\epsilon(X)\epsilon(Z)}[X,[Y,Z]]+ (-1)^{\epsilon(X)\epsilon(Y)}[Y,[Z,X]]+ (-1)^{\epsilon(Z)\epsilon(Y)}[Z,[X,Y]]=0, 
% \\ \forall X,Y,Z \in \mathfrak{g}, \nn
\end{align}
where $\epsilon(X)$ is the parity of $X$ and its value is determined depending on whether $X$ is Grassmann even or odd. It imposes a condition on $\lambda^a$s,
\be
(\lambda^a)^{\beta\gamma} (\lambda^a)^{\alpha}{}_{\delta}+ (\lambda^a)^{\alpha\gamma} (\lambda^a)^{\beta}{}_{\delta} = \frac{C_{\rho}}{d-1} (2 \eta^{\alpha\beta} \delta^{\gamma}_{\delta}-\eta^{\alpha\gamma} \delta^{\beta}_{\delta}-\eta^{\gamma\beta} \delta^{\alpha}_{\delta}).
\ee
The matrices $\eta^{\alpha\beta}$ are symmetric, i.e. $\eta^{\alpha\beta}=\eta^{\beta\alpha}$ while the matrices $\lambda^{\alpha\beta}$ are antisymmetric, i.e. $\lambda^{\alpha\beta}= -\lambda^{\beta\alpha}$.

A generic $\mathfrak{osp}(\cN|2)$ connection $\cA$ will be parameterized as
\begin{equation}
\cA = A^n L_n + B^a T^a + \psi^{\pm}_\alpha Q^\alpha_{\pm} = A + B + \Psi\,,
\end{equation}
and likewise for the barred sector.

The relation between the Chern-Simons formulation and the geometric formulation in terms of the dreibein $e$ and the (dualized) spin connection $\omega$ is obtained by identifying $sl(2)$ part of the super-connection as the dreibein and the spin connection, i.e. taking the $\mathfrak{sl}(2|\bR)$ connections $A, \Ab$ to be
\begin{equation}
e = \frac{\ell}{2}\left( A - \Ab \right)\,, \qquad \omega = \frac12 \left(A + \Ab \right)\,.
\end{equation}
The metric is constructed as
\begin{equation}\label{gdef}
g_{\mu\nu} = 2 \tr(e_{\mu} e_{\nu}) = \frac{\ell^2}{2} \tr \left( (A- \Ab)_{\mu} (A- \Ab)_{\nu}  \right)\,.
\end{equation}
All solutions of three dimensional gravity are locally gauge equivalent to each other, but differ up to boundary terms or global identifications. Therefore boundary conditions are crucial in three dimensional gravity, which we will now digress on at some length.

\subsubsection{Gauge fixing}\label{sec:gauge_fixing}
The importance of boundary conditions for three dimensional gravity stems from the fact that locally gauge equivalent solutions can differ by boundary terms. Imposing suitable boundary conditions in three dimensional gravity then specifies which bulk solutions are gauge inequivalent and lead to different boundary charges. Suitable in this context means: leading to integrable boundary charges. In addition, the boundary conditions should not be too strict, so that the boundary charges are non-trivial, but not too loose either, such that the boundary charges are finite.

In the Chern-Simons formulation of three dimensional gravity the boundary conditions on the gauge connection are most easily represented in a radial gauge. Let us suppose our manifold $\cM$ has the topology of a filled cylinder and can be equipped with a coordinate system $(t,\varphi,r)$. There is a spatial boundary $\partial \cM$ at $r \to \infty$. The radial gauge fixing is achieved by taking\footnote{For simplicity, here we work in one chiral sector of AdS$_3$ Chern-Simons gravity, unless otherwise specified.}
\begin{equation}
\partial_{\varphi} \cA_r = 0\,.
\end{equation} 
This implies that we can solve the Chern-Simons constraint $F_{r\varphi} = 0$ by writing 
\begin{equation}
\cA_r = b(r)^{-1} \partial_r b(r)\,, \qquad \cA_\varphi = b(r)^{-1} a_{\varphi}(t,\varphi) b(r)\,.
\end{equation}
Here $b(r)$ is a group element depending only on $r$. 

The time component of the Chern-Simons connection $\cA_t$ is a Lagrange multiplier for the constraint $F_{r\varphi} = 0$ and asking this constraint to be preserved under time evolution implies that we may write
\begin{equation}
\cA_t = b(r)^{-1} a_t (t,\varphi)b(r)\,.
\end{equation}
In this gauge, the boundary conditions are completely specified by fixing $a_{\varphi}(t,\varphi)$ and $a_t(t,\varphi)$. The functional variation of the asymptotic charges corresponding to an asymptotic symmetry transformation $\delta_{\ve}a_i = \partial_i \ve + [a_i,\ve \}$ \footnote{Here $[ ,\}$ is shorthand notation for the graded commutator.} are given by \cite{Banados:1998pi}
\begin{equation}\label{delQ}
\delta Q = -\frac{k}{2\pi} \oint \extd \varphi \; \Str \left( \ve \delta a_{\varphi} \right)\,.
\end{equation}
Hence, to specify suitable boundary conditions means in this context to find a form of $a_{\varphi}$ such that the above charges are finite and integrable for all $\ve$ which satisfy $ \partial_i \ve + [a_i,\ve \} = \delta_{\ve} a_\varphi $. An important aspect of this is to include a specification of state dependence in $\ve$ and in $a_\varphi$ in order to perform the functional integration of the charges.

It is now clear that $a_{\varphi}$ contains information on the asymptotic charges. On the other hand $a_t$ plays the role of chemical potential, or the sources. We may always write $a_t$ proportional to $\ve$, as the field equations $\partial_t a_\varphi - \partial_\varphi a_t + [a_t,a_\varphi\} =0$ will reduce to the statement that the time derivative of the charges $a_\varphi$ is determined by a symmetry transformation and hence weakly vanishes. The on-shell Chern-Simons action is proportional to $\int_{\partial\cM} \tr (a_t a_\varphi)$ and so by writing $a_t$ proportional to the gauge parameter $\ve$ one immediately sees that it becomes a chemical potential for the charges in $a_\varphi$.

Now we should specify the boundary conditions by specifying the form of $a_{\varphi}$. The gauge invariant observables in Chern-Simons theory are Wilson loops, or the holonomy of the connection around the $\varphi$-cycle. 
\begin{equation}
H_{\varphi} = \tr \left( \cP \exp \left( \oint a_{\varphi} \extd \varphi \right) \right)
\end{equation}
The holonomies, and therefore the distinct solutions, can be characterized by the conjugacy classes of the super-group $G$. In Section \ref{sec:bosonic_WZW}, we discuss how this can be useful in distinguishing different bulk solutions and classifying them as certain orbits of the asymptotic symmetry group in the bosonic example of Brown-Henneaux boundary conditions \cite{Brown:1986nw} for asymptotically local AdS$_3$ spacetimes.

\subsubsection{Supersymmetric Brown-Henneaux}\label{sec:susy-BH}

The supersymmetric extension of the Brown-Henneaux boundary conditions were obtained in \cite{Henneaux:1999ib}. In our conventions they are most easily represented by taking
\begin{equation}\label{conn-hwg}
	a_\vp = L_- + \cL(t,\vp) L_+ + \Psi_\alpha(t,\vp) Q^\alpha_+ + \cB^a(t,\vp)  T^a
\end{equation}
An asymptotic symmetry transformation $\delta a_\vp = \partial_\vp\Lambda + [a_\vp,\Lambda\}$ is parameterized by 
\begin{equation}\label{gauge-param-susy}
	\Lambda = \chi^n L_n + \ve^\pm_\alpha Q^\alpha_\pm + \omega^a T^a \,.
\end{equation}
Three components of $\Lambda$ can be solved for as
\begin{align}
	\chi^0 & = - \partial_\vp \chi^- \,, \qquad \chi^+ = - \frac12 \partial^2 \chi^-  + \cL \chi^- + \frac12 \Psi \cdot \ve^- \,, \\
	\ve^+_{\alpha} & = - \partial_\vp \ve^-_{\alpha} + \chi^- \Psi_{\alpha} + \cB^a \ve^-_{\alpha} \lambda^a \,,
\end{align}
where $\cdot$ means contraction by $\eta^{\alpha \beta}$.
The variation of the functions $\cB^a, \Psi_\alpha$ and $\cL$ are given by:
\begin{align}
	\delta \cB^a & = \partial_\vp \omega^a + f^{abc} B^b \omega^c - \frac{d-1}{2C_\rho} (\Psi \lambda^a \ve^-)\,,\label{delB} \\
	\delta \Psi_{\alpha} & = - \partial^2_\vp \ve^-_{\alpha} + \cL\, \ve^-_{\alpha} + \frac32 \partial_\vp \chi^- \Psi_{\alpha} + \chi^- \partial_\vp \Psi_{\alpha}  + \left(\partial_\vp \cB^a \ve^-_{\alpha} + 2 \cB^a \partial_\vp \ve^-_{\alpha} - \cB^a \cB^b \ve^-_{\alpha} \lambda^b \right)\lambda^a \nonumber \\
	& \qquad + \omega^a \Psi_{\alpha} \lambda^a - \cB^a \chi^- \Psi_{\alpha} \lambda^a \,. \label{delPsi} \\
	\delta \cL & = \chi^- \partial_\vp \cL + 2 \partial_\vp \chi^- \cL - \frac12 \partial^3_\vp \chi^- + \frac12 \partial_\vp \Psi \cdot \ve^- + \frac32 \Psi \cdot \partial_\vp \ve^- + \cB^a(\Psi \lambda^a \ve^-) \,. \label{delL}
\end{align}
The transformation rule for $\Psi$ contains a non-linear term $\sim \cB^a \Psi$ in the last line of \eqref{delPsi}. This term can be removed by taking the $R$-symmetry gauge parameter $\omega^a$ to be
\begin{equation}
\omega^a = \cB^a \chi^- + \Omega^a\,.
\end{equation}
This changes $\omega$ into $\Omega$ and removes the last term in \eqref{delPsi}, while it also changes \eqref{delB} to 
\begin{equation}
	\delta \cB^a = \partial_\vp(\cB^a \chi^-) + \partial_\vp \Omega^a + f^{abc} B^b \Omega^c - \frac{d-1}{2C_\rho} (\Psi \lambda^a \ve^-)\,,\label{delB2}
\end{equation}
It also has an effect on the asymptotic charges where it implements a Sugawara shift in the $\chi^-$ charge. The charges are
\begin{align}
	Q[\chi^-] & = -\frac{k}{2\pi} \int \extd \vp \; \chi^- \left(\cL + \frac{C_\rho}{d-1} \cB^a \cB^a \right) \equiv -\frac{k}{2\pi} \int \extd\vp \; \chi^-\hat{\cL} \,,\label{chargeL} \\
	Q[\ve^-] & = \frac{k}{2\pi} \int \extd\vp \; \ve^- \cdot \Psi \,, \\
	Q[\Omega] & = -\frac{k}{2\pi} \frac{2C_\rho}{d-1}\int \extd \vp \; \Omega^a \cB^a\,. 
\end{align}
The algebra of Fourier modes of the charges, defined as $L_n = Q[\chi^- = e^{in\vp}], G_n^\alpha = Q[\ve^-_\alpha = e^{in\vp}] $ and $T_n^a = Q[\Omega^a = e^{in\vp}]$, gives the generic $\cN$ superconformal algebra
\begin{subequations} \label{superconforml_alg}
\begin{align}
[L_m,L_n] & = (m-n)L_{m+n} + \frac{k}{2} m^3 \delta_{m+n,0}\,, \\
[L_m, G_n^\alpha] & = \left(\frac{m}{2}-n\right) G_{m+n}^\alpha \,, \\
[L_m, T_n^a] & = -n T_{m+n}^a \,, \\
\{G_m^{\alpha}, G_n^{\beta}\} & = \eta^{\alpha\beta} ( L_{m+n} + k m^2 \delta_{m+n,0} ) + i \frac{d-1}{2C_\rho} (m-n) (\lambda^a)^{\alpha\beta} T^a_{m+n} \\
& \qquad - \frac{k}{2} \left(\frac{d-1}{2k C_\rho} \right)^2 \left( \{\lambda ^a , \lambda^b\}^{\alpha\beta} + \frac{2C_\rho}{d-1} \eta^{\alpha\beta} \delta^{ab}\right)(T^a T^b)_{m+n} \,, \\
[G_m^\alpha, T_n^a]&  = i (\lambda^a)^{\alpha}{}_{\beta} G^\beta_{m+n} \,, \\
[T_m^a, T_n^b] & = i f^{abc} T_{m+n}^c + \frac{2C_\rho k}{d-1} m \delta^{ab} \delta_{m+n,0}\,.
\end{align}
\end{subequations}
where $(T^a T^b)_{m+n}$ are the Fourier modes of $(\frac{k}{2\pi} \frac{2C_\rho}{d-1})^2 B^a(\vp) B^b(\vp)$. This algebras agrees with \cite{Henneaux:1999ib} and corresponds to the non-linear superconformal algebras found in \cite{Knizhnik:1986wc,Bershadsky:1986ms,Fradkin:1992bz,Fradkin:1992km}. The non-linear terms in the supercharge anti-commutators are only present for $\cN>2$. For $\cN \leq 2$ the algebras are superconformal Lie algebras whose group is the centrally extended group of diffeomorphisms of the supercircle.

%%%%%%%%%%%%%%%%%%%%%%%%%%%%%%%%%%%%%%%%%%%%%%%%%%%%%%%%%%%%%%%%%%%%%%%%%%
%%%%%%%%%%%%%%%%%%%%%%%%%%%%%%%%%%%%%%%%%%%%%%%%%%%%%%%%%%%%%%%%%%%%%%%%%%

\subsection{$W_3$ higher spin gravity}

%In this note, we address the question of zero modes in the gravity theories  with a non-linear asymptotic symmetries. We will drive the boundary action through hamiltonian reduction and we claim that it is the geometric action on the symplectic leaves of the corresponding Poisson manifold. As an example of a geometric action on the symplectic leaves of a Poisson manifold, we consider $W_3$ gravity in $3$-dimensions as a simplest case where a non-linear algebra of charges appears.  

$W_3$ gravity in three dimensions can be considered as a higher spin theory of gravity for a spin-3 field and described by the Chern-Simons theory with the gauge group $SL(3,\mathbb{R})$. The Einstein Hilbert action is again written as the difference of two Chern-Simons actions as in \eqref{Seh} and \eqref{Scs},
where now $k$ is the level of the Kac-Moody $\mathfrak{sl}(3)$ algebra, the gauge connections $\cA,\bar{\cA}$ are $\mathfrak{g}=\mathfrak{sl}(3,\mathbb{R)}$-valued 1-form connections and the trace is over the $\mathfrak{sl}(3)$ Lie algebra. To define $\mathfrak{g}$ we use the root space gradation $\mathfrak{g}=\oplus \mathfrak{g}_{(i)}$ with
\bea
\mathfrak{g}_{(0)}&=&\mathfrak{g}_0 = \textrm{span}\{\theta^a | a=1,2,...,\textrm{rank}\,\mathfrak{g}\},\nonumber\\
\mathfrak{g}_{(j)}&=&\textrm{span}\{E^\alpha | [\theta^a,E^\alpha]= \alpha^a E^\alpha, a=1,2,...,\textrm{rank}\,\mathfrak{g} \},
\eea
where $\mathfrak{g}_0$ is the Cartan subalgebra of $\mathfrak{sl}(3)$ spanned by two elements $\theta^a$, $a=1,2$ and $E^\alpha$ are associated to the root $\alpha$ of the root space of $\mathfrak{sl}(3)$ Lie algebra. The Chevalley basis of $\mathfrak{sl}(3)$ are spanned by a set of 8 generators given by $\left\{E_{\pm}^{\alpha},\theta^a\right\}$, where $E_+^\alpha$, $E_-^\alpha$ are respectively raising operators corresponding to positive roots and lowering operators corresponding to negative roots, and $\theta^a$ span the Cartan subalgebra. We recall that the eigenvalues of the Cartan subalgebra are weights $\omega^a$ and the vector with components $\omega^a$ is a weight vector $\vec{\Omega}$, i.e. $\theta^a|\omega\rangle=\omega^a|\omega\rangle$ where $|\omega\rangle$ are eigenvectors of $\theta^a$. 

Let's consider the following matrix representation of $sl(3)$
\begin{align}\label{eq:chevalley_basis}
E_+^1 &= \begin{pmatrix} 0 & 1 & 0 \\ 0 & 0 & 0 \\ 0 & 0 & 0 \end{pmatrix}, \qquad~ E_+^2 = \begin{pmatrix} 0 & 0 & 0 \\ 0 & 0 & 1 \\ 0 & 0 & 0 \end{pmatrix}, \qquad E_+^3 = \begin{pmatrix} 0 & 0 & 1 \\ 0 & 0 & 0 \\ 0 & 0 & 0 \end{pmatrix},\nn\\
E_-^1 &= \begin{pmatrix} 0 & 0 & 0 \\ 1 & 0 & 0 \\ 0 & 0 & 0 \end{pmatrix}, \qquad E_-^2 = \begin{pmatrix} 0 & 0 & 0 \\ 0 & 0 & 0 \\ 0 & 1 & 0 \end{pmatrix}, \qquad E_-^3 = \begin{pmatrix} 0 & 0 & 0 \\ 0 & 0 & 0 \\ 1 & 0 & 0 \end{pmatrix},\nn\\
H_1 &= \begin{pmatrix} 1 & 0 & 0 \\ 0 & -1 & 0 \\ 0 & 0 & 0 \end{pmatrix}, \qquad H_2 = \begin{pmatrix} 0 & 0 & 0 \\ 0 & 1 & 0 \\ 0 & 0 & -1 \end{pmatrix},
\end{align}
and 
\begin{align}\label{eq:cartan_basis_sl3}
\theta^1 = \frac{H_1 + H_2}{\sqrt{2}},\qquad \qquad \theta^2 = \frac{H_1 -H_2}{\sqrt{6}}.
\end{align}
Using the matrix representation of $\mathfrak{sl}(3)$, we find three weight vectors of $\mathfrak{sl}(3)$
\be
\vec{\Omega}_1=\frac{1}{\sqrt{2}}\left(1,\frac{1}{\sqrt{3}}\right),\qquad
\vec{\Omega}_2=\frac{1}{\sqrt{2}}\left(0,-\frac{2}{\sqrt{3}}\right),\qquad
\vec{\Omega}_3=\frac{1}{\sqrt{2}}\left(-1,\frac{1}{\sqrt{3}}\right).
\ee
Roots are weights of the adjoint representation and root vectors $\vec{\alpha}_i$ can be defined as vectors connecting the weight vectors in the root lattice, then we can easily write the positive roots
\bea
\vec{\alpha}_1&=&\vec{\Omega}_1-\vec{\Omega}_2=\frac{1}{\sqrt{2}}\left(1,\sqrt{3}\right),\nonumber\\
\vec{\alpha}_2&=&\vec{\Omega}_2-\vec{\Omega}_3=\frac{1}{\sqrt{2}}\left(1,-\sqrt{3}\right),\nonumber\\
\vec{\alpha}_3&=&\vec{\Omega}_1-\vec{\Omega}_3=\left(\sqrt{2}, 0\right).
\eea
It is obvious that $\vec{\alpha}_1$ and $\vec{\alpha}_2$ are the simple roots of $\mathfrak{sl}(3)$, and $\vec{\alpha}_3=\vec{\alpha}_1+\vec{\alpha}_2$. This is simply understood from the commutator of the Cartan generators with raising and lowering operators as
\begin{align}
[\vec{\theta}, E^{i}_{\pm}]&=\pm \vec{\alpha}_i E^{i}_{\pm},\\
[E^{i}_{\pm}, E^{i}_{\mp}]&=\vec{\alpha}_i . \vec{\theta}\\
[E^{i}_{\pm}, E^{j}_{\pm}]&=\pm E^{i+j}_{\pm},
\end{align}
where $\vec{\theta} = (\theta_1,\theta_2)$. The rest of generators can be obtained by Serre relation or by the use of Jacobi Identity. Note that here ``." refers to the inner product on the two dimensional root vector space and moreover $E^{i}_{\pm} = 0$ for $i>3$.    

\subsubsection{Gauge fixing}
A general $\mathfrak{sl}(3)$ connection can be written as
\be
\cA = A_{\pm}^i E_{\pm}^i + \Theta^i H^i,
\ee
and likewise for $\bar{\mathcal{A}}$. Once again, the gauge inequivalent bulk solutions are specified by suitable boundary conditions. Following the discussion of section \ref{sec:gauge_fixing}, we can solve the Chern-Simons constraint in radial gauge for the connections 
\be
\mathcal{A}_r = b(r)^{-1}\partial_r b(r), \qquad \qquad \mathcal{A}_\vp = b(r)^{-1} a_{\vp}(t,\vp) b(r),
\ee
where $b(r)$ is a group element which only depends on $r$. Moreover, requiring that the Chern-Simons constraint remains unchanged under the time evolution leads to 
\be
\mathcal{A}_t = b(r)^{-1} a_{t}(t,\vp) b(r).
\ee
This means that the boundary conditions are completely determined by $a_\vp(t,\vp)$ and $a_t(t,\vp)$, see \cite{Campoleoni:2010zq, Balog:1990mu, Henneaux:2010xg} for more details.

The Brown-Henneaux boundary conditions for $W_3$ gravity in 3 dimensions depends on the way that the $SL(2)$ subgroup is embedded in $SL(3)$. We will explore this in more detail later in section \ref{sec:HWG} where we discuss the Hamiltonian reduction of $SL(3,\mathbb{R)}$ WZW action.

%It has been shown that the Poisson bracket of its asymptotic charges (on the boundary of AdS$_3$), form a $W_3$ algebra as the asymptotic symmetry of the theory hence the name $W_3$ gravity. There are two different $sl(3)$ representation depending on the embedding of its $sl(2)$ subalgebra, $sl(2)$ principal embedding and diagonal embedding. The choice of embedding affects the highest weight gauge fixing of the connection on the boundary and in turn it may affect the form of the reduced action on the boundary. We will discuss both embedding in the section \ref{sec:HWG}.

%However, we can write down the WZW action with $SL(3,\mathbb{R})$ without resorting to the choice of embedding. The zero modes of boundary currents are determined by the holonomy and one can choose the monodromy group. We focus on the hyperbolic holonomy. The other classes follow from the similar analysis.

%%%%%%%%%%%%%%%%%%%%%%%%%%%%%%%
%%%%%%%%%%%%%%%%%%%%%%%%%%%%%%%
\section{Hamiltonian reduction of the super Chern-Simons action}\label{sec:reduction}

In this section we will discuss the reduction of three dimensional super-Chern-Simons action to a two dimensional field theory. We first reduce three dimensional AdS supergravity to a sum of two chiral super Wess-Zumino-Witten models and then impose the boundary conditions as constraints on the WZW super-currents. Furthermore, we will discuss the effect of different boundary conditions and boundary terms on the two dimensional boundary action, paying special attention to non-trivial holonomies in the bulk. The analysis is kept general for extended supergravity with any number of supersymmetries. For $\mathcal{N}\leq 2$, we will show how the boundary action is related to the geometric action on the coadjoint orbit of the super-Virasoro group. For larger supersymmetries ($\mathcal{N} >2$), we compute the Schwarzian action through the reduction, and suggest a form for the geometric action on the symplectic leaves of corresponding super-Virasoro group manifold.

%In this section we will reduce the Chern-Simons action to a two dimensional field theory by first reducing three dimensional AdS gravity to a sum of two chiral Wess-Zumino-Witten models and then imposing the boundary conditions as constraints on the WZW currents. We will discuss the effect of different boundary conditions and boundary terms on the two dimensional boundary action, paying special attention to non-trivial holonomies in the bulk. We will show how the boundary action is related to the geometric action on the coadjoint orbit of the Virasoro group. The holonomy of the Chern-Simons connection turns out to correspond to the constant representative on the coadjoint orbit.

The first steps of the reduction follows along the lines of \cite{Coussaert:1995zp,Henneaux:1999ib} (see also \cite{Donnay:2016iyk} for a recent review). What is new in our approach is that we will not try to combine the two chiral sectors of the theory, but instead keep them disconnected. We also discuss the zero modes of the fields and allow for non-trivial holonomies of the Chern-Simons connection. 

We start with the Hamiltonian decomposition of the action \eqref{Seh}. Let us focus on one chiral sector ($S_{\textsc{cs}}[A]$) for now, as the barred sector follows similarly. The Hamiltonian form of the action \eqref{Scs} is defined on a manifold $\cM$ which we will take to be the disk $D$ times a time direction. The boundary $S^1$ of the disk is at $r\to \infty$ and we use an orientation $ dt \wedge d\vp \wedge d r$. The Hamiltonian action is 
\begin{equation}\label{ScsHam}
S[\cA] = \frac{k}{4\pi} \int_{\cM} dt  d\varphi dr\, \Str \left(\cA_{r} \dot{\cA}_{\varphi} - \cA_{\varphi} \dot{\cA}_{r} + 2 \cA_{t} \mathcal{F}_{\varphi r} \right) + I_{\Sigma} \,,
\end{equation}
with
\begin{equation}
\label{Fphir}
\mathcal{F}_{r \vp} = \partial_{r} \cA_\vp - \partial_\vp \cA_r + [\cA_r, \cA_\vp]\,,
\end{equation}
and $I_{\Sigma}$ is a boundary term adapted to the boundary conditions under consideration.
This boundary term is added to the action to ensure a well-defined variational principle. The variation of the bulk action gives
\begin{equation}\label{bdyterm}
\delta S[\cA] = \ldots -  \frac{k}{2\pi} \int_{\partial \cM} dt d\varphi \, \Str \cA_t \delta \cA_{\varphi} = \ldots -  \frac{k}{2\pi} \int_{\partial \cM} dt d\varphi \, \Str \; a_t \delta a_{\varphi}  \,. 
\end{equation}
Here the ellipses stand for terms proportional to the bulk equations of motion. The boundary term $I_{\Sigma}$ should be such that its variation cancels the term in \eqref{bdyterm}. We should therefore impose boundary conditions such that this expression is integrable and finite.  We discuss the boundary conditions in details later in Section \ref{sec:Ham}.

The action \eqref{ScsHam} consists out of two parts. There is a symplectic term and the boundary term. The latter is responsible for the Hamiltonian of the boundary theory, while the symplectic part reduces to the geometric action on the super-Virasoro coadjoint orbit. In the non-supersymmetric theory, it will reduce to the geometric action on the Virasoro coadjoint orbits of \cite{Alekseev:1988ce}. Let us discuss the reduction for both these terms in succession.

\subsection{The symplectic term}\label{sec:WZW}
Focusing first on the bulk part of one chiral sector of the action, we see that $\cA_t$ is a Lagrange multiplier for the constraint $\mathcal{F}_{\varphi r}=0$.
In general, the spatial sections can have a non-trivial topology, which implies a non-trivial holonomy along non-contractible cycles in the manifold. One can take this into account in the following way: 

One demands the group elements $G$ to be periodic; as a result of which the holonomy appears explicitly in the local solution of the constraint  $\mathcal{F}_{\varphi r}=0$, 
\begin{equation}\label{explicitAphi}
\cA_{\vp} = G^{-1}(\partial_\vp + K(t)) G\,, \qquad G(\vp+2\pi) = G(\vp).
\end{equation}
Here, $K(t)$ is a Lie algebra valued function of time that parametrizes the holonomy\footnote{This choice is indeed a restriction we impose at this point, as it is only possible to eliminate the $\vp$-dependence in $K(t)$ for simply connected groups.}.
Note that for periodic group elements $G$, the integral of $G^{-1}\partial_\vp G$ over the loop vanishes.

The action with explicit holonomy can be obtained by substituting \eqref{explicitAphi} into \eqref{ScsHam}. 
The result is (formula (A.7) of  \cite{Henneaux:1999ib})
\begin{align}\label{Sexpl}
S_{\rm SCS}[G, K(t)] = & + \frac{k}{4\pi} \int_{\cM} d^3x \; \Str \left( \partial_r (  G^{-1} \partial_\vp  G  G^{-1} \partial_t  G ) \right) + \frac{k}{12\pi} \int_{\cM}  \; \Str ( G^{-1} d  G)^3 \\
& + \frac{k}{4\pi} \int_{\cM} d^3x \; \Str \left(2 \partial_r ( G^{-1} K \partial_t G ) - \partial_t (G^{-1} K \partial_r G) \right)\nonumber + I_{\Sigma}\,,
\end{align}
up to a total $\vp$-derivative which is dropped due to the periodicity of $G$ in $\vp$. In addition, we have also dropped boundary contributions at the time boundaries, and we will continue to do so in the following, up to the point where we discuss them systematically.

In Section \ref{sec:reduction-SWZW}, we continue by representing any group element with the use of the Gauss decomposition of the group. It enables us to define Maurer-Cartan one-forms, $G^{-1} dG$, and to further implement constraints from the boundary conditions, i.e. $a_\vp = G^{-1} \partial_\vp G$, in terms of the fields parameterizing the group. But first, we will discuss the Hamiltonian coming from the boundary term in \eqref{ScsHam}.

\subsection{The Hamiltonian}\label{sec:Ham}
Imposing different boundary conditions can lead to different boundary Hamiltonians. Following \eqref{bdyterm}, the boundary term should be taken such that its variation is 
\begin{equation}
\delta I_{\Sigma} = \frac{k}{2\pi} \int dt d\vp \; \Str\left(\cA_t \delta \cA_\vp \right)\,.
\end{equation}
The precise form of boundary action depends on the topology of spatial surfaces. In the following we focus on a case where the spatial hypersurface has annulus topology $S^1 \times [0,1]$.

In that case, we impose the boundary condition $\cA_- = 0 $ of \cite{Coussaert:1995zp} at the outer boundary. 
We also choose the Hamiltonians on the respective boundaries to have the same (positive) sign for definiteness, which one could interpret as having time evolution on both sides of the black hole run in the same direction. This is achieved by taking  $\cA_+ = 0$ at the inner boundary. These choices lead to the boundary terms (see section 2.3 of \cite{Henneaux:2019sjx}):
\begin{equation}
\label{Hamiltonian}
I_{\Sigma_{i,o}} = -\frac{k}{4\pi} \int_{\Sigma_{i,o}} dt d\vp \, \Str \cA_\vp^2\,.
\end{equation}

The boundary term in the action is thus seen as the time integral of the Noether charge of an asymptotic symmetry transformation and hence the resulting action will remain invariant under the asymptotic symmetries.
\begin{equation}\label{qbdy}
I_{\Sigma} = I_{\Sigma_i} + I_{\Sigma_o} = -\frac{k}{4\pi} \int dt H_{\rm bdy} =  -\int dt\, Q[ \Lambda = \cA_t ],
\end{equation}
where $\Lambda$ is a Lie algebra valued function of the gauge parameters. If we choose $\chi^- =1$ and $\ve^-=0 = \Omega^a$, then the asymptotic symmetry corresponds to time translations. The Noether charge for this would make a suitable Hamiltonian and it's given by
\begin{equation}\label{bdyham}
I_{\Sigma} = \frac{k}{2\pi} \int \extd t \extd\vp \; \hat{\cL}
\end{equation}
with $\hat{\cL}$ defined in \eqref{chargeL}.

For the annulus topology, the action includes a contribution from each boundary, coupled through the holonomy parameterized by $K$.
\begin{align}\label{WZW2bdy}
S_{\rm CS}[h,l, K(t)]  = & +  \frac{k}{4\pi} \int_{\Sigma_o} dt d\vp \; \Str \left( h^{-1} \partial_\vp  h  h^{-1} \partial_-  h  + 2 h^{-1} K \partial_- h   - K^2 \right)  \\
& - \frac{k}{4\pi} \int_{\Sigma_i} dt d\vp \; \Str \left( l^{-1} \partial_\vp  l  l^{-1} \partial_+  l  + 2 l^{-1} K \partial_+ l   + K^2 \right)  \nonumber + I_{WZ}[G] \,, \nonumber
\end{align}
where:
\be
h= G(t, r=r^{\textrm{outer}}, \varphi), \qquad l = G(t, r=r^{\textrm{inner}}, \varphi)\,,
\ee
and hereon we use the notation $\partial_{\mp} = \partial_t \mp \partial_{\vp}$.

The Wess-Zumino term
\begin{equation}
I_{WZ}[G] =  \frac{k}{12\pi} \int_{\cM}  \; \Str ( G^{-1} d  G)^3  \,,
\end{equation}
can be written as a total derivative and hence it also only depends on the boundary values of the group element $G$.

The action \eqref{WZW2bdy} is invariant under the gauge symmetry $G \rightarrow \omega(t) G$ and $K \rightarrow \omega(t) K \omega^{-1}(t)$, which implies in terms of the boundary fields,
\be \label{gaugetrafo}
h \rightarrow \omega(t) h, \qquad l \rightarrow \omega(t) l, \qquad K \rightarrow \omega(t) K \omega^{-1}(t)\,.
\ee
This gauge invariance results from the redundancy of the parametrization of the group element $G$ \cite{Elitzur:1989nr} and can  straightforwardly be verified in the above action.

\subsection{On the holonomy}\label{sec:reduction-SWZW}
In this section, we continue the discussion of Section \ref{sec:WZW} by explicitly computing the reduction for a super-Chern-Simons action with $\mathcal{N}$ extended supersymmetry and the gauge group $OSp(\cN|2)$. However before doing so, we need to clarify few important aspects of the presence of holonomy. As described in \cite{Henneaux:2019sjx}, the inequivalent physical contribution of holonomy $K$ is classified by its conjugacy classes, and moreover, conjugacy classes are constant in time.

%\subsubsection{Discussion on holonomy and supergroup}

Due to the theorem 3.9 of \cite{Berezin87} (see also \cite{kobayashi1988}), the holonomy group of $OSp(\cN|2)$ can be always put in a block diagonalized form. This means that one can always make the holonomy to be in the even subgroup of the supergroup. Another way to see this is by looking at the classification of coadjoint orbits of super Virasoro group \cite{Yang1991}. One notices that the coadjoint orbits that one obtains by considering the holonomy in the odd subspace of the superspace are not new ones and one could equivalently obtain them by setting the holonomy to be in the even subgroups.

In the case of $\mathcal{N}=1$ the choice of holonomy is insensitive to the supersymmetry. This is because for $\mathcal{N}=1$ there is no $R$-symmetry and as a result the holonomy is classified by the conjugacy classes of $SL(2)$, similar to the non-supersymmetric case. However, for $\mathcal{N}\geq 2$ there is an $SO(\cN)$ $R$-symmetry and the holonomy must be classified by the conjugacy classes of the even subgroup of $OSp(\cN|2)$.
 
Another remark on the choice of holonomy conjugacy classes is that we only focus on the hyperbolic holonomy class for the rest of this paper. This class of holonomies includes BTZ solutions and therefore is important in the study of two-sided eternal BTZ black holes which is the primary (but not the only) concern of this paper. For the practical reasons and for the sake of simplicity we only focus on the hyperbolic class. The analysis of the other classes is similar to the hyperbolic case. For such an analysis in the bosonic case see \cite{Henneaux:2019sjx}.

%[COMMENT: I was not able to find any description or classification for the conjugacy classes of even subgroup of $OSp(N|2)$ in the literature. Since I know how it is for $\mathcal{N}=1$, I write down the complete analysis for $\cN=1$ below.]

\subsection{$\cN=1$ super Chern-Simons}\label{sec:N1superCS} 

We will now continue to discuss the reduction to the boundary action for $\cN = 1$ AdS$_3$ supergravity, focusing on the case where the holonomy is in the hyperbolic conjugacy class of $SL(2)$.

\subsubsection{Hyperbolic Holonomy}

Using the Gauss decomposition, we can parameterize any element of $OSp(1|2)$ as 
\begin{align}
h = G[r=r_o, t,\vp] &= e^{Y L_- + \psi^- Q_-} e^{\Phi L_0} e^{X L_+ + \psi^+ Q_+} \equiv E^- E^0 E^+\, \label{Gauss_decomp_out_N1}\\
l = G[r=r_i, t,\vp] &= e^{U L_+ + \chi^+ Q_+} e^{\tilde{\Phi} L_0} e^{V L_- + \chi^- Q_-} \equiv \tilde{E}^+ \tilde{E}^0 \tilde{E}^-\, \label{Gauss_decomp_in_N1}
\end{align}
where $\Phi, X,Y, \psi^{\pm}$ are fields at outer boundary and $\tilde{\Phi}, U,V, \chi^{\pm}$ are fields defined at inner boundary and all are depending on spacetime coordinates $(t,\vp)$\footnote{We emphasize that the parametrization \eqref{Gauss_decomp_in_N1} is only at the inner boundary. One can see it as a field redefinition of the fields which appear in a similar parametrization as \eqref{Gauss_decomp_out_N1}.}. 
They are the pullback of the $r$-dependent $G$ to the corresponding boundaries.

Inserting  \eqref{Gauss_decomp_out_N1} and \eqref{Gauss_decomp_in_N1} inside the action \eqref{WZW2bdy}, upon using the algebra $\mathfrak{osp}(1|2,\mathbb{R})$ given in \eqref{ospN} and
\be
\Str(L_+ L_-) =1,\qquad \Str(L_0 L_0) = \frac12 , \qquad
\Str(Q_- Q_+) = - \Str(Q_+ Q_-) = 2,
\ee
the action now reads off as
\be
S = S_o - S_i +S_{hol},
\ee 
where\footnote{Here and in the rest of the paper we use prime as a shorthand notation for $\partial_\vp$.}
\begin{align}\label{Sdecomposed_N1}
S_o &=  \frac{k}{4\pi} \int_{\Sigma_o} \extd t \extd \vp \; \; \Big[ \frac12  \Phi'  \partial_{-} \Phi +  2 e^{\Phi} (\partial_{-} X - \psi^+ \partial_{-} \psi^+)(Y' - \psi^- \psi^-{}') - 4 e^{\Phi/2} \psi^-{}'  \partial_{-} \psi^+   \Big], \nonumber \\
S_i &=  \frac{k}{4\pi} \int_{\Sigma_i} \extd t \extd \vp \; \; \Big[ \frac12  \tilde{\Phi}'   \partial_{+}\tilde{\Phi} +  2 e^{-\tilde{\Phi}} (\partial_{+} V + \chi^- \partial_{+} \chi^-)(U' + \chi^+ \chi^+{}') - 4 e^{-\tilde{\Phi}/2} \partial_{+} \chi^- \chi^+{}'    \Big],
\end{align} 
and
\begin{align}
S_{hol} = \frac{k}{4\pi} \int \extd t \extd \vp \; \; \Big[ & k_0 \bigg(\partial_{-}\Phi - \partial_{+} \tilde{\Phi}+  2 e^{\Phi/2} \psi^- \partial_{-}\psi^+ - 2 e^{-\tilde{\Phi}/2} \chi^+ \partial_{+} \chi^-\nn\\
&-2 e^{\Phi} Y( \partial_{-} X - \psi^+ \partial_{-}\psi^+)-2 e^{-\tilde{\Phi}} U(\partial_{+} V + \chi^- \partial_{+} \chi^-)\bigg) - k_0^2 \Big].
\end{align}

The Lagrangian is invariant (up to total derivatives) under the residual gauge symmetry,
\begin{subequations}
\begin{align}
\Phi \to \hat \Phi & = \Phi + \lambda^0 \qquad\qquad \tilde{\Phi} \to \hat{ \tilde{\Phi}} = \tilde{\Phi} + \lambda^0 , \\
Y \to \hat Y & =  Y  e^{- \lambda^0}  \qquad\qquad\, U \to \hat{U} =  U  e^{\lambda^0} ,\\
X \to \hat X & = X  \qquad\qquad\qquad V \to \hat{V} = V , \\
\psi^- \to \hat{\psi}^- & = \psi^- e^{- \lambda^0/2} \quad\quad \chi^- \to \hat{\chi}^- = \chi^- ,  \\
\psi^+ \to \hat{\psi}^+ & = \psi^+  \qquad\qquad\,~ \chi^+ \to \hat{\chi}^+ = \chi^+ e^{- \lambda^0/2} , \\
k_0 \to \hat k_0 & = k_0 \, .
\end{align}
\end{subequations}
A convenient rewriting of the action is
\begin{equation}
S_{\rm CS}[k_0, Y,\Phi,X,\psi^{\pm},V,\tilde{\Phi},U,\chi^{\pm}] = S_{\rm bdy}^{\Sigma_o}[k_0, Y, \Phi, X,\psi^{\pm}] - S_{\rm bdy}^{\Sigma_i}[k_0, V, \tilde{\Phi}, U,\chi^{\pm}] \,,
\end{equation}
with
\bea
S_{\rm bdy}^{\Sigma_o} &= \frac{k}{4\pi} \int dt d\vp \; \Big[\frac{1}{2}\partial_- \Phi (\Phi' + 2 k_0) + 2 e^{\Phi} (\partial_- X - \psi^+ \partial_-\psi^+) (Y' - \psi^- {\psi^-}' - k_0 Y) \nonumber\\
&+ 4 e^{\Phi/2} \partial_- \psi^+ ({\psi^-}' -\frac{k_0}{2} \psi^-)  - \frac{1}{2} k_0^2 \Big]\,,\nonumber\\
S_{\rm bdy}^{\Sigma_i} &= \frac{k}{4\pi} \int dt d\vp \; \Big[\frac{1}{2} \partial_+ \tilde{\Phi} ({\tilde{\Phi}}' + 2 k_0) + 2 e^{-\tilde{\Phi}} (\partial_+ V + \chi^- \partial_+ \chi^-) (U' + \chi^+ {\chi^+}' + k_0 U) \nonumber\\
&- 4 e^{-\tilde{\Phi}/2} \partial_+ \chi^- ({\chi^+}' + \frac{k_0}{2} \chi^+)  + \frac{1}{2} k_0^2 \Big]\,,\nonumber\\\label{Sbdy}
\eea

\subsubsection{Boundary conditions}
We are now at the stage where we can impose the reduction conditions on the super Chern-Simons connection that expresses the asymptotic behaviour of extended (super) $AdS_3$ \cite{Henneaux:1999ib}. We consider explicitly one asymptotic boundary only (the outer boundary).   Similar considerations apply to the inner boundary. The only difference in their treatment is the choice of $OSp(N|1,\mathbb{R})$ representation at each boundary. While the boundary conditions on the fields at outer boundary are in accordance with highest-weight representation, those on the fields at inner boundary are in accordance with the lowest-weight representation. As shown in \cite{Coussaert:1995zp} and discussed in \cite{Henneaux:1999ib, Henneaux:2019sjx} the boundary conditions on the fields at $r=r_2 \equiv r_o$ are \begin{equation}
\cA_r = 0, \qquad \cA_\vp = L_- + \cL(t,\vp) L_+ + \Psi^+(t,\vp) Q_+.
\label{eq:DSReduction_N1}
\end{equation}
Similarly, the boundary conditions on the fields at $r=r_1 \equiv r_i$ are \begin{equation}
\cA_r = 0, \qquad \cA_\vp = L_+ + \cM(t,\vp) L_- + \Psi^-(t,\vp) Q_-.
\label{eq:DSReduction_lowest_N1}
\end{equation}
In terms of the field appearing in the Gauss decomposition, this gives the conditions
\begin{align}
\begin{array}{c||c}
\textrm{On Outer Boundary} & \textrm{On Inner Boundary}\\
\hline\hline
~ & ~ \\
	e^{\Phi} (Y' -\psi^-{\psi^-}' -k_0 Y)  = 1  &  e^{-\tilde{\Phi}} (U' + \chi^+ {\chi^+}' +k_0 U) = 1  \\
~ & ~\\	
	\Phi' + k_0  = 2 X 	&  \tilde{\Phi}' + k_0 = - 2 V  \\
~ & ~ \\	
    \psi^+ + e^{\Phi/2} ({\psi^-}' -\frac{k_0}{2} \psi^-)  = 0  &  
    \chi^- + e^{-\tilde{\Phi}/2} ({\chi^+}' + \frac{k_0}{2} \psi^+)  = 0    
\end{array}
\end{align}
Inserting these conditions in the action \eqref{Sbdy}, we find
\bea
S_{\rm bdy}^{\Sigma_o} &= \frac{k}{4\pi} \int dt d\vp \; \Big[\frac{1}{2}\partial_- \Phi (\Phi' + 2 k_0) + 2 \psi^+ \partial_-\psi^+ - \frac{1}{2} k_0^2 \Big]\,,\nonumber\\
S_{\rm bdy}^{\Sigma_i} &= \frac{k}{4\pi} \int dt d\vp \; \Big[\frac{1}{2} \partial_+ \tilde{\Phi} ({\tilde{\Phi}}' + 2 k_0) - 2 \chi^- \partial_+ \chi^- + \frac{1}{2} k_0^2 \Big]\,.\label{S_reduced}
\eea
Putting these together, the final form of action is written 
\bea\label{S_reduced_full}
S = \frac{k}{4\pi} \int dt d\vp \; \Big[\frac{1}{2}\partial_- \Phi \Phi' - \frac{1}{2} \partial_+ \tilde{\Phi} {\tilde{\Phi}}' + k_0 (\partial_- \Phi - \partial_+ \tilde{\Phi}) + 2 (\psi^+ \partial_-\psi^+ + \chi^- \partial_+ \chi^-)- k_0^2 \Big]\,.\nn\\
\eea
This action describes the dynamic of two chiral bosons (one at each boundary) linked through the holonomy and two free chiral fermions (one at each boundary). This action is exactly the one described in \cite{Henneaux:2019sjx} with the exception of the free chiral fermions present at each boundary and similarly enjoys invariance under a redundant gauge symmetry
\bea
\Phi \rightarrow \Phi + \epsilon(t), \qquad \tilde{\Phi} \rightarrow \tilde{\Phi} + \epsilon(t), \qquad k_0 \rightarrow k_0 \,,\nonumber\\
\psi^+ \rightarrow \psi^+, \qquad \chi^- \rightarrow \chi^-\,.
 \label{eq:GaugeSymmAnnulus_N1}
\eea

It is of no surprise that fermions remained uncoupled as the holonomy belongs only to the diagonal subgroup which only allows bosonic degrees of freedom to couple. In the next section, we should see an extension of this to include $R$-symmetry.

Before proceeding any further lets make a comment on the other conjugacy classes of $SL(2)$; elliptic and parabolic conjugacy classes. As we just pointed out since for $\cN =1$, holonomy only couples to the bosonic fields, the treatment of elliptic and parabolic cases are similar to \cite{Henneaux:2019sjx} in each case. The only difference is the presence of free chiral fermions in the action which will appear as in the action \eqref{S_reduced_full}.

\subsection{$\cN=2$ super Chern-Simons}\label{sec:N2superCS}

The case of $\cN=2$ seems fairly easy to deal with. In fact, one can write down the $S_o$ and $S_i$ for a general $\cN >2$. The difficulty lies in what is the contribution of $S_{hol}$ which depends on the choice of holonomy. 
Our guideline motivated by the theorem on supergroup diagonalization \cite{Berezin87, kobayashi1988} is that the holonomy can be always put in the form where it belongs to the even subgroup of the supergroup. For $\cN=2$ this simplifies drastically, since there is only one $R$-symmetry generator and the holonomy will be given by a direct sum of the $\cN=1$ holonomy (i.e. it belongs to one of $SL(2,\mathbb{R})$ conjugacy classes) and the $\mathfrak{so}(2)$ Lie algebra of the $\cN=2$ $R$-symmetry subgroup\footnote{This choice of holonomy can be also shown to be the case for a constant orbit representative, when one considers the effect of the holonomy implicitly through non-periodicity of the fields.}. This boils down to three distinct classes: 
\begin{itemize}
\item \textbf{Hyperbolic} holonomy described by hyperbolic conjugacy class of $\mathfrak{sl}(2) \oplus \mathfrak{so}(2)$. Explicitly, the holonomy group element is $H = \exp{K}$ with $K(t) = k_0(t) L_0 + k_r(t) T$. All elements conjugate to $K$ belong to this class.
\item \textbf{Elliptic} holonomy described by the elliptic conjugacy class of $\mathfrak{sl}(2)$, or explicitly: all elements conjugate to $K(t) = \frac{1}{2}k_e(t) (L_- - L_+)$.
\item \textbf{Parabolic} holonomy, which is described by the parabolic conjugacy class of $\mathfrak{sl}(2)$, or explicitly: all elements conjugate to $K(t) = k_p(t) L_+$.
\end{itemize}

Before discussing the details for $\cN=2$, let us first write down the outer and inner contributions to the action $S_o$ and $S_i$, for general $\cN$.

Once again we are only discussing the hyperbolic holonomy conjugacy classes which contains BTZ back holes. The Gauss decomposition is a natural parametrization for the hyperbolic holonomy. It occurs that the $R$-symmetry generator for $\cN=2$ is also diagonal, which makes the choice of Gauss decomposition much more suitable for this case.

Using the Gauss decomposition, we can parameterize any element of $OSp(\cN|2)$ as 
\begin{align}
h = G[r=r_o, t,\vp] &= e^{Y L_- + \psi^-_{\alpha} Q^{\alpha}_-} e^{\Phi L_0 + {\rm{i}}\, C^a T^a} e^{X L_+ + \psi^+_{\alpha} Q^{\alpha}_+} \equiv E^- E^0 E^+\, \label{Gauss_decomp_out}\\
l = G[r=r_i, t,\vp] &= e^{U L_+ + \chi^+_{\alpha} Q^{\alpha}_+} e^{\tilde{\Phi} L_0 + {\rm{i}}\, \tilde{C}^a T^a} e^{V L_- + \chi^-_{\alpha} Q^{\alpha}_-} \equiv \tilde{E}^+ \tilde{E}^0 \tilde{E}^-\, \label{Gauss_decomp_in}
\end{align}
where $\Phi, X,Y, C^a,\psi^{\pm}_{\alpha}$ are fields at outer boundary and $\tilde{\Phi}, U,V, \tilde{C}^a, \chi^{\pm}_{\alpha}$ are fields defined at inner boundary and all are depending on spacetime coordinates $(t,\vp)$. They are the pullback of the $r$-dependent $G$ to the corresponding boundaries.

Inserting  \eqref{Gauss_decomp_out} and \eqref{Gauss_decomp_in} inside the action \eqref{WZW2bdy} and using the $\mathfrak{osp}(N|2,\mathbb{R})$ algebra given in \eqref{ospN} and
\begin{align}
\Str(L_+ L_-) &=1,\qquad\qquad\quad~~ \Str(L_0 L_0) = \frac12,\nn\\
\Str(T^aT^b) &= \frac{2C_\rho}{d-1}\delta^{ab},\qquad \Str(Q^\alpha_- Q^\beta_+) = - \Str(Q^\alpha_+ Q^\beta_-) = \eta^{\alpha\beta},
\end{align}
the action now reads:
\be
S = S_o - S_i +S_{hol},
\ee 
where
\begin{align}\label{Sdecomposed}
S_o =  \frac{k}{4\pi} \int_{\Sigma_o} \extd t \extd \vp \; \; \Big[ &\frac12  \Phi'  \partial_{-} \Phi +  2 e^{\Phi} (\partial_{-} X - \psi^+_{\alpha} \frac{\eta^{\alpha\beta}}{2} \partial_{-} \psi^+_{\beta})(Y' - \psi^-_{\alpha} \frac{\eta^{\alpha\beta}}{2}\psi_{\beta}^-{}')\nn\\
&- 2 e^{\Phi/2} \psi_{\beta}^-{}' \mathfrak{m}^{\beta\alpha} \partial_{-} \psi^+_{\alpha}   \Big] + S_o[u], \nonumber\\
S_i =  \frac{k}{4\pi} \int_{\Sigma_i} \extd t \extd \vp \; \; \Big[ &\frac12  \tilde{\Phi}'   \partial_{+}\tilde{\Phi} +  2 e^{-\tilde{\Phi}} (\partial_{+} V + \chi^-_{\alpha} \frac{\eta^{\alpha\beta}}{2} \partial_{+} \chi^-_{\beta})(U' + \chi^+_{\alpha} \frac{\eta^{\alpha\beta}}{2} \chi_{\beta}^+{}')\nn\\
&+ 2 e^{-\tilde{\Phi}/2} \chi_{\beta}^+{}' \tilde{\mathfrak{m}}^{\beta\alpha} \partial_{+} \chi^-_{\alpha}  \Big]
+ S_i[w].
\end{align}
Here $\mathfrak{m}\indices{^\alpha_\beta}= e^{\textrm{i}\, C^a (\lambda^a)\indices{^\alpha_\beta}}$, $\tilde{\mathfrak{m}}\indices{^\alpha_\beta}= e^{\textrm{i}\, \tilde{C}^a (\lambda^a)\indices{^\alpha_\beta}}$, $\mathfrak{m}^{\alpha\beta} = \mathfrak{m}\indices{^\alpha_{\gamma}} \eta^{\gamma\beta}$, and $\tilde{\mathfrak{m}}^{\alpha\beta} = \tilde{\mathfrak{m}}\indices{^\alpha_{\gamma}} \eta^{\gamma\beta}$, with 
$\lambda^a$ the basis of the $\mathfrak{so}(\cN)$ algebra under which fermions transform, and
\begin{align}\label{sbdy_u}
S_o[u] &=  \frac{k}{4\pi} \int_{\Sigma_o} dt d\vp \; \Str \left( u^{-1} \partial_\vp  u  u^{-1} \partial_-  u \right) + I_{WZ}[U]|_{\Sigma_o}  \\
S_i[u] &= \frac{k}{4\pi} \int_{\Sigma_i} dt d\vp \; \Str \left( w^{-1} \partial_\vp  w  w^{-1} \partial_+  w   \right) + I_{WZ}[W]|_{\Sigma_i} \,. \nonumber
\end{align}
Here, once again, $u(t,\vp)= e^{\textrm{i}\, C^a(t,\vp) T^a}$, and $w(t,\vp)= e^{\textrm{i}\, \tilde{C}^a(t,\vp) T^a}$ are the $r$-independent pullback of $U(t,\vp ,r)= e^{\textrm{i}\, C^a(t,\vp ,r) T^a}$ and $W(t,\vp ,r)= e^{\textrm{i}\, \tilde{C}^a(t,\vp ,r) T^a}$, respectively.

\subsubsection{Hyperbolic holonomy}
As explained in the previous section, we consider that $K(t) = k_0 L_0 + k_r T$. For such a holonomy, the contribution of holonomy in the action is given by
\be
S_{hol} = S_{hol}^{k_0} + S_{hol}^{k_r}
\ee
with
\begin{align}
S_{hol}^{k_0} = \frac{k}{4\pi} \int \extd t \extd \vp \; \; \Big[ & k_0 \bigg(\partial_{-}\Phi - \partial_{+}\tilde{\Phi}+ e^{\Phi/2} (\psi^- \cdot \mathfrak{m}\cdot\partial_{-}\psi^+) - e^{-\tilde{\Phi}/2} (\chi^+\cdot\tilde{\mathfrak{m}}\cdot\partial_{+} \chi^-)\nn\\
&-2 e^{\Phi} Y( \partial_{-} X - \frac{1}{2} \psi^+ \cdot \partial_{-}\psi^+)-2 e^{-\tilde{\Phi}} U(\partial_{+} V + \frac{1}{2}\chi^- \cdot \partial_{+} \chi^-)
\bigg) - k_0^2 \Big],
\end{align}
and
\begin{align}
S_{hol}^{k_r} = \frac{k}{4\pi} \int \extd t \extd \vp \; \; \Big[ & k_r \left(-i\left(\partial_{-}C - \partial_{+}\tilde{C}\right)- e^{\Phi} (\psi^-\cdot\lambda\cdot\psi^-)( \partial_{-} X - \frac{1}{2}\psi^+ \cdot\partial_{-}\psi^+)\right. \nonumber\\
& + \left. e^{-\tilde{\Phi}} (\chi^+\cdot\lambda\cdot\chi^+)(\partial_{+} V + \frac{1}{2}\chi^- \cdot\partial_{+} \chi^-) + 2 e^{\phi/2} (\psi^-\cdot\lambda\cdot\mathfrak{m}\cdot\partial_{-}\psi^+)\right.\nonumber\\
& - \left. 2 e^{-\tilde{\phi}/2} (\chi^+\cdot\lambda\cdot\tilde{\mathfrak{m}}\cdot\partial_{+}\chi^-)\right) + k_r^2 \Big].
\end{align}

Here we have used a shorthand notation ``$\cdot$" to avoid writing all spinorial indices explicitly. As an example, it reads as $\psi^+ \cdot\partial_{-}\psi^+ = \psi^+ \cdot \eta \cdot \partial_{-}\psi^+ = \psi^+_{\alpha} \eta^{\alpha\beta}\partial_{-}\psi^+_{\beta}$ and $\psi^- \cdot\mathfrak{m}\cdot\partial_{-}\psi^+ = \psi^-_{\alpha} \mathfrak{m}^{\alpha\beta} \partial_{-}\psi^+_{\beta}$. Notice that here $\lambda$ (and consequently $\mathfrak{m}$) is given by the $\mathfrak{osp}(N|2)$ representation of appendix \ref{app:conv} and we have set $\frac{2C_{\rho}}{d-1} = -\frac{1}{2}$ accordingly.

The full $\cN=2$ super Chern-Simons action can then be written as
\begin{equation}\label{Sbdyn=2}
S_{\rm CS}[k_0,K_r, Y,\Phi,X,\psi^{\pm}_{\alpha},V,\tilde{\Phi},U,\chi^{\pm}_{\alpha}] = S_{\rm bdy}^{\Sigma_o}[k_0,k_r, Y, \Phi, X,\psi^{\pm}_{\alpha}] - S_{\rm bdy}^{\Sigma_i}[k_0,k_r, V, \tilde{\Phi}, U,\chi^{\pm}_\alpha] \,,
\end{equation}
with
\begin{align}
S_{\rm bdy}^{\Sigma_o} = \frac{k}{4\pi} \int dt d\vp \; \Big[ &\frac{1}{2}\partial_- \Phi (\Phi' + 2 k_0) + \frac{1}{2}\partial_- C (C' - 2i k_r)\nonumber\\
& + 2 e^{\Phi} (\partial_- X - \frac{1}{2}\psi^+ \cdot \partial_-\psi^+) (Y' - \frac{1}{2}\psi^-\cdot {\psi^-}' - k_0 Y -\frac{1}{2}k_r (\psi^-\cdot\lambda\cdot\psi^-)) \nonumber\\
& - 2 e^{\Phi/2} ({\psi^-}' -\frac{k_0}{2} \psi^- - k_r \psi^-\cdot\lambda)\cdot\mathfrak{m}\cdot\partial_- \psi^+   - \frac{1}{2} (k_0^2-k_r^2) \Big]\label{Sbdyn=2out}
\end{align}
and
\begin{align}
S_{\rm bdy}^{\Sigma_i} = \frac{k}{4\pi} \int dt d\vp \; \Big[ &\frac{1}{2}\partial_+ \tilde{\Phi} (\tilde{\Phi}' + 2 k_0) + \frac{1}{2}\partial_+ \tilde{C} (\tilde{C}' - 2i k_r)\nonumber\\
& + 2 e^{-\tilde{\Phi}} (\partial_+ V +\frac{1}{2} \chi^- \cdot \partial_+ \chi^-) (U' + \frac{1}{2}\chi^+ \cdot {\chi^+}' + k_0 U -\frac{1}{2}k_r (\chi^+\cdot\lambda\cdot\chi^+)) \nonumber\\
& + 2 e^{-\tilde{\Phi}/2} ({\chi^+}' +\frac{k_0}{2} \chi^+ + k_r \chi^+\cdot\lambda)\cdot\tilde{\mathfrak{m}}\cdot\partial_+ \chi^-   + \frac{1}{2} (k_0^2-k_r^2) \Big].\label{Sbdyn=2in}
\end{align}

\subsubsection{Boundary condition}

Next, we impose the boundary conditions on the $\cN=2$ super Chern-Simons connection, corresponding to asymptotically AdS$_3$ spacetimes \cite{Henneaux:1999ib}. The boundary conditions on the fields at outer boundary are in accordance with highest-weight representation, those on the fields at inner boundary are in accordance with the lowest-weight representation. As shown in \cite{Coussaert:1995zp} and discussed in \cite{Henneaux:1999ib, Henneaux:2019sjx}, the boundary conditions on the fields at $r=r_2 \equiv r_o$ are \begin{equation}
\cA_r = 0, \qquad \cA_\vp = L_- + \cL(t,\vp) L_+ + \Psi^+_{\alpha}(t,\vp) Q_+^{\alpha} + \cB(t,\vp) T.
\label{eq:DSReduction}
\end{equation}
Similarly, the boundary conditions on the fields at $r=r_1 \equiv r_i$ are \begin{equation}
\cA_r = 0, \qquad \cA_\vp = L_+ + \cM(t,\vp) L_- + \Psi^-_{\alpha}(t,\vp) Q_-^{\alpha} + \cR(t,\vp) T.
\label{eq:DSReduction_lowest}
\end{equation}
In terms of the field appearing in the Gauss decomposition, this gives the conditions
\begin{align}
{\footnotesize{
\begin{array}{c||c}
\textrm{On Outer Boundary} & \textrm{On Inner Boundary}\\
\hline\hline
~ & ~ \\
	e^{\Phi} [Y' -\frac{1}{2}\psi^-\cdot{\psi^-}' -k_0 Y -\frac{1}{2}k_r (\psi^-\cdot\lambda\cdot\psi^-)]  = 1  &  e^{-\tilde{\Phi}} [U' + \frac{1}{2}\chi^+\cdot{\chi^+}' + k_0 U -\frac{1}{2}k_r (\chi^+\cdot\lambda\cdot\chi^+)] = 1  \\
~ & ~\\	
	\Phi' + k_0  = 2 X 	&  \tilde{\Phi}' + k_0 = - 2 V  \\
~ & ~ \\	
    \psi^+_{\alpha} + e^{\Phi/2} [{\psi^-}' -\frac{k_0}{2} \psi^- - k_r (\psi^-\cdot\lambda)]_{\beta} \mathfrak{m}\indices{^\beta_\alpha}  = 0  &  
    \chi^-_{\alpha} + e^{-\tilde{\Phi}/2} [{\chi^+}' + \frac{k_0}{2} \psi^+ + k_r (\chi^+\cdot\lambda)]_{\beta} \tilde{\mathfrak{m}}\indices{^\beta_\alpha}  = 0    
\end{array}
}}
\nonumber\\
\end{align}
Inserting these conditions in the action \eqref{Sbdyn=2}, we find
\begin{align}
S_{\rm bdy}^{\Sigma_o} = \frac{k}{4\pi} \int dt d\vp \; \Big[ &\frac{1}{2}\partial_- \Phi (\Phi' + 2 k_0) + \frac{1}{2}\partial_- C (C' - 2i k_r)+ 2 \psi^+ \cdot(\mathfrak{m} -\frac{1}{2}\eta)\cdot \partial_-\psi^+ \nn\\
&- \frac{1}{2} (k_0^2-k_r^2) \Big]\,,\nonumber\\
S_{\rm bdy}^{\Sigma_i} = \frac{k}{4\pi} \int dt d\vp \; \Big[ &\frac{1}{2}\partial_+ \tilde{\Phi} (\tilde{\Phi}' + 2 k_0) + \frac{1}{2}\partial_+ \tilde{C} (\tilde{C}' - 2i k_r) - 2 \chi^- \cdot(\tilde{\mathfrak{m}}-\frac{1}{2}\eta)\cdot \partial_+ \chi^- \nn\\
&+ \frac{1}{2} (k_0^2-k_r^2) \Big]\,.\label{S_reducedn=2}
\end{align}
Putting these together, the final form of action is written with pairs $(\Phi,\tilde{\Phi})$ and $(C,\tilde{C})$ of chiral bosons linked through the holonomy and two pairs of chiral fermions with a dilaton coupling to the $C$ field 
\begin{align}
S = \frac{k}{4\pi} \int dt d\vp \; \Big[ &\frac{1}{2}\partial_- \Phi \Phi'-\frac{1}{2}\partial_+ \tilde{\Phi} \tilde{\Phi}'  + k_0 (\partial_- \Phi-\partial_+ \tilde{\Phi})\nn\\
&+ \frac{1}{2}\partial_- C C' - \frac{1}{2}\partial_+ \tilde{C} \tilde{C}' -i k_r(\partial_- C-\partial_+ \tilde{C})\nn\\
&+2 \psi^+ \cdot(\mathfrak{m} -\frac{1}{2}\eta)\cdot \partial_-\psi^+ + 2 \chi^- \cdot(\tilde{\mathfrak{m}}-\frac{1}{2}\eta)\cdot \partial_+ \chi^-\nn\\
&-(k_0^2-k_r^2) \Big]\,.\label{eq:S_complete_any_N}
\end{align}
For $\cN=2$, see appendix \eqref{app:conv}, we have
\be
\mathfrak{m}-\frac{1}{2}\eta = \begin{pmatrix}
 0 & 2 e^{-\frac{iC}{2}}-1 \\ 2 e^{\frac{iC}{2}}-1 & 0
\end{pmatrix}.
\ee

This action enjoys invariance under a redundant gauge symmetry
\begin{align}
\Phi &\rightarrow \Phi + \epsilon_1(t), \qquad \tilde{\Phi} \rightarrow \tilde{\Phi} + \epsilon_1(t)\,,\nonumber\\
C &\rightarrow C + \epsilon_2(t) , \qquad  \tilde{C} \rightarrow \tilde{C}+ \epsilon_2(t) \,,\nonumber\\
k_0 &\rightarrow k_0,\qquad \qquad ~~~ k_r \rightarrow k_r \,,\nonumber\\ 
\psi^+_{\alpha} &\rightarrow \psi^+_{\alpha}, \qquad \quad ~~~~ \chi^-_{\alpha} \rightarrow \chi^-_{\alpha}\,.
 \label{eq:GaugeSymmAnnulus}
\end{align}

The action \eqref{eq:S_complete_any_N} is one of the main result of this paper. Even though this action is written for $\cN =2$ supergravity yet it contains general feature that holds for any $\cN$. The holonomy independent part of the action always will be written in this form considering a generalization of $R$-symmetry term. For larger $\cN$ the holonomy dependent part is very much related to the chosen conjugacy class of holonomy.

Another observation regarding the action \eqref{eq:S_complete_any_N} is that it is equivalent to the sum of two chiral actions of \cite{Henneaux:1999ib} when the holonomies are set to zero. In another words, this is a generalization of \cite{Henneaux:1999ib} by including the zero modes and holonomies. In \cite{Henneaux:1999ib} it was shown that for the bosonic part and ignoring zero modes and holonomies, this is equivalent to the Liouville action. A detail analysis of such a relation in the case of super Liouville action and in the presence of zero modes and holonomies will be discussed elsewhere \cite{AR2023}.

Last comment regarding the action \eqref{eq:S_complete_any_N} is that this is indeed the action on the super Virasoro coadjoint orbit. We will come back to this point in Section \ref{sec:N=2} where we explicitly show that the action \eqref{eq:S_complete_any_N} is equivalent to the geometric action on the coadjoint orbit of the group of reparameterizations of $\cN =2$ supercircle. 

\subsection{Comments on General $\cN$}\label{sec:generalN}

For $\mathcal{N}>2$, the $R$-symmetry group is larger, and there are many more conjugacy classes. The number of spinors will also be much more, leading to higher than quadratic orders of interactions. Although the basic idea of holonomy is the same, the detailed analysis will be much more complicated.

However, for the case when the holonomy is given by $K(t) = k_0 L_0 + k_r T$, where $T$ is one of the Cartan elements of the $R$-symmetry algebra, the action \eqref{S_reducedn=2} is indeed the reduced Hamiltonian action. 

Moreover, for a constant orbit representative, we are able to write down the complete action for any $\cN$, by considering the holonomy as non-periodicity in the fields. However, this comes at a price. The drawback of treating the holonomies as non-periodicity in the fields is that the holonomy is no longer a dynamical field. In the quantized theory of gravity one expects to have all the possible solutions such that the dynamics of the theory allows changing between the solutions. In the geometric picture all these solutions are determined by different orbits and therefore a sensible classical action requires to have the orbit representatives (holonomies) as dynamical fields, such that in the partition function one can sum over all the possible solutions of them.

All this makes us believe that writing the action in the presence of a dynamical holonomy requires a case-by-case study that we do not intend to pursue here.

%%%%%%%%%%%%%%%%%%%%%%%%%%%%%%%%%%%%%%%%%%%%%%%%%%%%%%%%%%%%%%%%%%%%%%%%%%%%%%%%%%%%%%%%%%%%%%%%%%%%%%%%%%%%%%%%%%%%%%%%%%%%%%%%%%%%%%%%%%%%%%%%%%%%%
\section{Hamiltonian reduction of $W_3$ higher Chern-Simons action}\label{sec:Higher-Spin}

As we have repeatedly mentioned, the goal of this paper is to provide a boundary action through Hamiltonian reduction of various extension of Chern-Simons theory where the algebra of asymptotic charges is non-linear while emphasizing the role of zero modes and holonomies. Until now, we discussed the supersymmetric extension of Chern-Simons theory of gravity where the non-linear terms will appear in $\cN >2$ supergravity models. Another place where a non-linear algebra of charges will appear is in higher spin extension of Chern-Simons gravity \cite{Campoleoni:2010zq,Henneaux:2010xg}. For a spin-$N$ field the asymptotic charges form a $W_N$ algebra. This algebra is non-linear for $N>2$, however we have much less fields compared to supergravity and a simpler classification of holonomy conjugacy classes. In this section, we will focus on a Chern-Simons theory for a spin-3 fields with $W_3$ asymptotic symmetry algebra and we write down the boundary action via Hamiltonian reduction. We will show this for different representations of $SL(3)$ and we claim that the boundary action in each case is the geometric action on the symplectic leaves of $W_3$ manifold.

\subsection{Chern-Simons action in the presence of zero modes}
In this section, we follow closely the discussion of section \ref{sec:reduction}, however now the gauge fields are $\mathfrak{sl}(3,\mathbb{R})$ valued. As explained, the Chern-Simons action \eqref{Scs} can be written as
\begin{equation}\label{ScsHam_W3}
S[\cA] = \frac{k}{4\pi} \int_{\cM} dt  d\varphi dr\, \tr \left( \cA_{\varphi} \dot{\cA}_{r} - \cA_{r} \dot{\cA}_{\varphi} + 2 \cA_{t} \mathcal{F}_{r \varphi} \right) + I_{\Sigma_{i}} + I_{\Sigma_{o}}\,,
\end{equation}
with
\begin{equation}
\label{Fphir2}
\mathcal{F}_{r \vp} = \partial_{r} \cA_\vp - \partial_\vp \cA_r + [\cA_r, \cA_\vp]\,.
\end{equation}
and $I_{\Sigma_{i,o}}$ are boundary terms adapted to the boundary conditions under consideration.

Imposing the boundary condition $\cA_- =0$ at outer boundary and $\cA_+ =0$ at inner boundary, we have for the boundary Hamiltonian actions $I_{\Sigma_{i,o}}$
\begin{equation}
I_{\Sigma_{i,o}} = -\frac{k}{4\pi} \int_{\Sigma_{i,o}} dt d\vp \, \tr \cA_\vp^2\,.
\end{equation}
The field $\cA_t$ is a Lagrange multiplier for the constraint $\mathcal{F}_{r\vp}=0$ which can be solved for if
\begin{align}
\cA_\vp = G^{-1} \left( \partial_\vp + K(t) \right) G, \qquad \qquad \cA_r =  G^{-1}\partial_r G.
\end{align}
Here $G$ is an $SL(3,\mathbb{R})$ element and $K(t)$ is the Lie-algebra valued holonomy. For our purpose of dealing with hyperbolic holonomy, it is suitable to choose a representative in the Gauss decomposition of $SL(3,\mathbb{R})$, i.e.
\begin{align}
 G = e^{E_-^\alpha} e^{\theta^a} e^{E_+^\alpha},
\end{align}
where $\{E_{\pm}^\alpha, \theta^a\}$, $\alpha=1,2,3$, $a=1,2$ are generators of $\mathfrak{sl}(3,\mathbb{R})$, which form the Chevalley basis of $\mathfrak{sl}(3,\mathbb{R})$. A matrix representation in terms of this basis is given by \eqref{eq:chevalley_basis} and \eqref{eq:cartan_basis_sl3}.

As we have discussed in the $\mathfrak{sl}(2)$ case, the zero modes are related to the non-trivial holonomies in the bulk, which are characterized by both the conjugacy classes of $\mathfrak{sl}(3)$, and the classification of symplectic leaves \cite{cmp/1104249646,Bajnok:2000nb}. In the case of $SL(3)$, there are seven different conjugacy classes; two hyperbolic, one elliptic, three parabolic and the exceptional conjugacy classes. A complete list of conjugacy classes of the $SL(3)$ group based on its possible isotropy subgroups can be found in \cite{Bajnok:2000nb}. 

Once again we focus in our discussion only on holonomies in the non-degenerate hyperbolic conjugacy class since BTZ black holes have hyperbolic holonomies. This holonomy class can be represented by diagonalizable matrices with three distinct real eigenvalues which are determined with two parameters\footnote{The dimension of isotropy subgroups of $SL(3)$ is $2,4$ or $8$ which means the holonomies can be parametrized with these number of parameters \cite{Bajnok:2000nb}.}. These two parameters can be completely fixed by asymptotic symmetry transformations which set them to the asymptotic charges $\cL$ and $\mathcal{W}$. This is similar to the Chern-Simons theory with Virasoro symmetry of charges. In that case the holonomies are determined by isotropy groups of $SL(2)$ group, which leads to three conjugacy classes. These classes are determined only with one parameter which upon fixing by an appropriate Virasoro transformation is set to the Virasoro charge $\cL$. Indeed for the hyperbolic holonomy with a fixed orbit representative $b_0$, the zero modes of $\cL$ is determined by $b_0$ which is related to the only parameter in the holonomy matrix $k_0$. For the details see section \ref{sec:bosonic_WZW}. In the case of $SL(3)$ group, zero modes of $\cL$ and $\mathcal{W}$ are determined by the intersections of two symplectic leaves with the representatives $b_0$ and $b_1$. These representatives are related to two holonomy parameters $k_0$ and $k_1$.

\subsubsection{Hyperbolic holonomy}
When dealing with hyperbolic holonomies, a suitable parametrization of group elements $G$ is given by the Gauss decomposition, where we have:
\begin{align}\label{Gauss_decom_sl3}
G = e^{\gamma_3 E_-^3} e^{\gamma_2 E_-^2 + \gamma_1 E_-^1} e^{\frac{1}{\sqrt{2}}\phi_1 \theta_1 + \sqrt{\frac{3}{2}}\phi_2 \theta_2} e^{\xi_1 E_+^1 + \xi_2 E_+^2} e^{\xi_3 E_+^3}.
\end{align}
Here $\left\{E_{\pm}^{\alpha},\theta^a\right\}$ are Chevalley basis of $\mathfrak{sl}(3)$ and are defined in \eqref{eq:chevalley_basis} and \eqref{eq:cartan_basis_sl3}. 

In terms of these group elements, the action can be written as
\begin{align}\label{eq:actionSL3}
S_{\rm CS}[G, K(t)] = & + \frac{k}{4\pi} \int_{\cM} d^3x \; \tr \left( \partial_r (  G^{-1} \partial_\vp  G  G^{-1} \partial_t  G ) \right) + \frac{k}{12\pi} \int_{\cM}  \; \tr ( G^{-1} d  G)^3 \\
& + \frac{k}{4\pi} \int_{\cM} d^3x \; \tr \left(2 \partial_r ( G^{-1} K \partial_t G ) - \partial_t (G^{-1} K \partial_r G) \right)\nonumber + I_{\Sigma_{i}} + I_{\Sigma_{o}}\,.
\end{align}
Let's consider an annulus geometry with two boundaries $r^o=r^{\textrm{outer}}$ and $r^i = r^{\textrm{inner}}$ such that
\be
h= G(t, r=r^{\textrm{outer}}, \varphi), \qquad l = G(t, r=r^{\textrm{inner}}, \varphi)\,.
\ee

Similarly to our previous discussions, we would like to impose lowest weight gauge condition on the inner boundary and it is more practical to use a field redefinition on the inner boundary in order to parametrize the group elements $l$ as
\begin{align}\label{Gauss_decom_sl3_inner}
l = e^{\tilde{\xi}_3 E_+^3} e^{\tilde{\xi}_2 E_+^2 + \tilde{\xi}_1 E_+^1} e^{\frac{1}{\sqrt{2}}\tilde{\phi}_1 \theta_1 + \sqrt{\frac{3}{2}}\tilde{\phi}_2 \theta_2} e^{\tilde{\gamma}_1 E_-^1 + \tilde{\gamma}_2 E_-^2} e^{\tilde{\gamma}_3 E_-^3}.
\end{align}

Then, the action on the $r$-boundaries reduces to
\begin{align}
S_{\rm CS}[h,l, K(t)]  = & +  \frac{k}{4\pi} \int_{\Sigma_o} dt d\vp \; \tr \left( h^{-1} \partial_\vp  h  h^{-1} \partial_-  h  + 2 h^{-1} K \partial_- h   - K^2 \right)  \\
& - \frac{k}{4\pi} \int_{\Sigma_i} dt d\vp \; \tr \left( l^{-1} \partial_\vp  l  l^{-1} \partial_+  l  + 2 l^{-1} K \partial_+ l   + K^2 \right)  \nonumber + I_{WZ}[G] \,, \nonumber
\end{align}
As usual, the Wess-Zumino term
\begin{equation}
I_{WZ}[G] =  \frac{k}{12\pi} \int_{\cM}  \; \tr ( G^{-1} d  G)^3  \,,
\end{equation}
can be written as a total derivative and hence it also only depends on the boundary values of the group element $G$.

For the non-degenerate hyperbolic holonomy, we have
\be
K(t) = k_0(t) (H_1 + H_2) + k_1(t) H_2.
\ee
This is the hyperbolic class with three distinct real eigenvalues parameterized by two arbitrary functions of $t$.

Now, given the reparametrization \eqref{Gauss_decom_sl3}, the action \eqref{eq:actionSL3} takes the form

\begin{align}\label{eq:outer_boundary}
S_o = \frac{k}{4\pi} \int_{\Sigma_o} dt\, & d\vp \bigg(2 e^{\frac{\phi_1 +3 \phi_2}{2}} \left[\gamma_1' - (k_0 -k_1)\gamma_1\right]\, \partial_- \xi_1\nn\\
&+ 2 e^{\frac{\phi_1 - 3 \phi_2}{2}} \left[\gamma_2' - (k_0+2k_1)\gamma_2\right]\, \partial_- \xi_2  \nn\\
&+\frac{1}{2} e^{\phi_1} F(\gamma_1,\gamma_2,\gamma_3;k_0,k_1)\,\left(\xi_2\partial_- \xi_1 - \xi_1\partial_- \xi_2 -2 \partial_- \xi_3 \right)\nn\\
&+\frac{1}{2}\left[\left(\phi_1' + 4 k_0 +2 k_1\right) \partial_-\phi_1 + 3 \left(\phi_2' -2 k_1\right) \partial_-\phi_2 \right]\nn\\
&- 2\left(k_0^2 +k_0 k_1 + k_1^2 \right) \bigg),
\end{align}
where
\begin{equation}
    F(\gamma_1,\gamma_2,\gamma_3;k_0,k_1)=\gamma_2\gamma_1' -\gamma_1\gamma_2' -2\gamma_3' + k_1(2\gamma_3+ 3 \gamma_1 \gamma_2) + 4 k_0 \gamma_3.
\end{equation}
The action on the inner boundary $S_i$ is a completely similar expression to the action on the outer boundary except written in terms of the fields on inner boundary $\tilde{\gamma}, \tilde\phi, \tilde{\xi}$ instead of $\gamma,\phi,\xi$ respectively.

\subsection{Different $SL(3)$ representations and highest weight gauge condition}\label{sec:HWG}

The finite dimensional representations of $SL(3)$ can be obtained from those of $SL(2)$ by considering different non-isomorphic $SL(2)$ subgroups embedded in $SL(3)$. This can be understood for example from the decomposition of finite dimensional representation of $SL(3)$ group into the eigenspaces of the vector space formed from the Cartan elements of $\mathfrak{sl}(3)$ algebra \cite{Fulton2004}. These representations come with a unique highest weight. For $SL(3)$ there are two distinct representations: principal and diagonal $SL(2)$ embeddings. In each case, the root space decomposition differ which in turn leads to a different (but unique in each representation) highest weight. In the following we discuss each representation and the corresponding boundary actions. The boundary actions will be different given that the highest weight gauge condition for the reduction is different in each representation.

\subsubsection{Boundary Condition and Hamiltonian Reduction}

In this section we follow closely the reduction procedure discussed in sections \ref{sec:N1superCS} and \ref{sec:N2superCS}. Therefore, we impose an appropriate boundary condition at each boundary. As before, we consider that the angular component of $\mathfrak{sl}(3)$ Chern-Simons connection to be fixed by the highest weight or the lowest weight representations on the outer or inner boundaries respectively. This is consistent with the asymptotic behaviour of AdS$_3$.

\subsubsection{Principal Embedding} 

With the principal embedding, one can choose the basis such that the $\mathfrak{sl}(3)$ algebra is written as
\begin{align}\label{sl(3)_principal}
[L_m,L_n] &= (m-n) L_{m+n},\\
[L_m,W_i] &= (2m-i) W_{m+i},\\
[W_i,W_j] &= -\frac{1}{3} (i - j)(2 i^2 + 2 j^2 - ij-8)L_{i+j},\label{eq:WW_commutator} 
\end{align} 
where $L_m$'s with $m=\pm 1,0$ are the $\mathfrak{sl}(2)$ generators and $W_i$'s with $i=\pm 2,\pm 1,0$ are the rest of $\mathfrak{sl}(3)$ generators. One can obtain these generators directly from the Chevalley basis \cite{Li:2015osa} as
\begin{align}
L_0 &\equiv \frac{1}{2} \sum_{i=1}^{2} (c_i H^i),\\
L_{\pm 1}  &\equiv \pm \sum_{i=1}^{2} 2^{\mp\frac{1}{2}}\sqrt{c_i} E_{\mp}^i,\\
W_{i} &\equiv (-1)^{2-i} \frac{(2+i)!}{12} \left(\textrm{adj}_{L_{-1}}\right)^{2-i} \left(L_{+1}\right)^2,
\end{align}
with $c_i = 2 \sum_j (C^{-1})_{ij}$ where $C$ is the Cartan matrix of $\mathfrak{sl}(3)$ algebra
\be
C = \begin{pmatrix} 2 & -1 \\ -1 & 2 \end{pmatrix}.
\ee

An explicit basis for this algebra is then given by \cite{Campoleoni:2010zq}

\begin{align}
L_{-1} &= \begin{pmatrix} 0 & -2 & 0 \\ 0 & 0 & -2 \\ 0 & 0 & 0 \end{pmatrix}, \qquad~ L_{0} = \begin{pmatrix} 1 & 0 & 0 \\ 0 & 0 & 0 \\ 0 & 0 & -1 \end{pmatrix}, \qquad L_{+1} = \begin{pmatrix} 0 & 0 & 0 \\ 1 & 0 & 0 \\ 0 & 1 & 0 \end{pmatrix},\nn\\
W_{-1} &= \begin{pmatrix} 0 & -2 & 0 \\ 0 & 0 & 2 \\ 0 & 0 & 0 \end{pmatrix}, \qquad W_{0} = \frac{2}{3}\begin{pmatrix} 1 & 0 & 0 \\ 0 & -2 & 0 \\ 0 & 0 & 1 \end{pmatrix}, \qquad W_{+1} = \begin{pmatrix} 0 & 0 & 0 \\ 1 & 0 & 0 \\ 0 & -1 & 0 \end{pmatrix},\nn\\
W_{-2} &= \begin{pmatrix} 0 & 0 & 8 \\ 0 & 0 & 0 \\ 0 & 0 & 0 \end{pmatrix}, \qquad W_{+2} = \begin{pmatrix} 0 & 0 & 0 \\ 0 & 0 & 0 \\ 2 & 0 & 0 \end{pmatrix}.\nn
\end{align}

The highest and lowest weight gauge conditions in terms of the $\mathfrak{sl}(2)$-principal embedding basis takes the following form \cite{Campoleoni:2010zq,Henneaux:2010xg}
\begin{align}
\cA_r &= 0, \qquad \cA_\vp = L_{+1} + \cL(t,\vp) L_{-1} + \mathcal{W}(t,\vp) W_{-2} \qquad \textrm{at outer boundary}\\
\cA_r &= 0, \qquad \cA_\vp = L_{-1} + \cL(t,\vp) L_{+1} + \mathcal{W}(t,\vp) W_{+2} \qquad \textrm{at inner boundary}.
\end{align}
At the outer boundary, this results in the following conditions
\begin{align}
e^{\frac{\phi_1 +3 \phi_2}{2}} \left[\gamma_1' - (k_0 -k_1)\gamma_1\right] &=1,\\
 e^{\frac{\phi_1 - 3 \phi_2}{2}} \left[\gamma_2' - (k_0+2k_1)\gamma_2\right] &=1,\\
F(\gamma_1,\gamma_2,\gamma_3;k_0,k_1) &= 0,\\
\xi_2 - (k_0 + k_1) - \frac{1}{2} \left(\phi_1' - \phi_2'\right) &= 0,\\
\xi_1 - k_0 - \frac{1}{2} \left(\phi_1' + \phi_2'\right) &= 0,
\end{align}
together with the expressions for $\cL$ and $\mathcal{W}$

\begin{align}
\cL &= -\frac{1}{4}\left(\xi_1^2 +3 \xi_2^2 -2 \xi_1 \xi_2 +2 \xi_3 + \xi_1' +\xi_2'\right),\\
\mathcal{W} &= \frac{1}{8}\left(\xi_3' + \frac{1}{2}(\xi_2\xi_1'-\xi_2'\xi_1)+\xi_2(\xi_1^2-2\xi_1\xi_2+2\xi_3)\right).
\end{align}

Inserting these condition in the action, we obtain the reduced boundary action on the outer boundary
\begin{align}\label{eq:action_W3_out}
S_{\rm bdy}^{\Sigma_o}[k_0, k_1,\phi_1,\phi_2] =  \frac{k}{4\pi} \int dt d\vp \; \bigg( &\frac{1}{2}\partial_- \phi_1 (\phi'_1 + 4 k_0 + 2 k_1) + \frac{3}{2}\partial_- \phi_2 (\phi'_2 -2 k_1)\nn\\
&- 2(k_0^2 + k_0 k_1 + k_1^2) \bigg).
\end{align}
Following the same steps at the inner boundary, the complete boundary action can be written as
\begin{align}
S_{\rm bdy}[k_0, k_1,\phi_1,\phi_2,\tilde{\phi}_1,\tilde{\phi}_2] =  \frac{k}{4\pi} \int dt d\vp \, &\bigg(\frac{1}{2}\partial_- \phi_1 \,\phi'_1 - \frac{1}{2}\partial_+ \tilde{\phi}_1 \, \tilde{\phi}'_1 
 +\frac{3}{2}\partial_- \phi_2 \, \phi'_2 - \frac{3}{2}\partial_+ \tilde{\phi}_2 \, \tilde{\phi}'_2 \nn\\
& + (2k_0 + k_1)(\partial_- \phi_1- \partial_+ \tilde{\phi}_1) - 3 k_1(\partial_- \phi_2 - \partial_+ \tilde{\phi}_2)\nn\\ 
& - 4(k_0^2 + k_0 k_1 + k_1^2) \bigg),
\end{align}
where $\psi_1$ and $\psi_2$ are the fields on the inner boundary.

This action has a redundant gauge symmetry
\begin{align}\label{eq:gauge_red_W3}
\phi_1 &\rightarrow \phi_1 + \epsilon_1(t), \qquad \tilde{\phi}_1 \rightarrow \tilde{\phi}_1 + \epsilon_1(t)\,,\nonumber\\
\phi_2 &\rightarrow \phi_2 + \epsilon_2(t), \qquad \tilde{\phi}_2 \rightarrow \tilde{\phi}_2 + \epsilon_2(t)\,,\nonumber\\
k_0 &\rightarrow k_0,\qquad \qquad ~~~~ k_1 \rightarrow k_1 \,.
\end{align}

In order to write the action \eqref{eq:action_W3_out}, we have eliminated $\gamma_1$ and $\gamma_2$ while keeping $\phi_1$ and $\phi_2$. One can instead keep $\gamma_1$, $\gamma_2$ in which case one can write the action as the $\mathfrak{sl}(3)$ generalization of the Schwarzian action upon the following field redefinitions:
\begin{align}
\gamma_1 &= e^{-(k_0-k_1)(f(t,\vp)-\vp)},\label{eq:field_redef_W3_1}\\
\gamma_2 &= e^{-(k_0+2k_1)(g(t,\vp)-\vp)}\label{eq:field_redef_W3_2}.
\end{align}
At the outer boundary, the action is now
\begin{align}\label{eq:action_schw-W3}
S &= \frac{k}{4\pi} \int dt d\vp \, \left[ -\frac{4}{3}\left( \left[\frac{\partial_- f''}{f'}-\left(\frac{3}{2}\frac{f''}{f'}+\frac{1}{4}\frac{g''}{g'}\right)\frac{\partial_- f'}{f'} \right]\right.\right. \nn\\
&~~~~~~~~~~~~~~~~~~~~~~~~+  \left.\left.\left[\frac{\partial_- g''}{g'}-\left(\frac{3}{2}\frac{g''}{g'}+\frac{1}{4}\frac{f''}{f'}\right)\frac{\partial_- g'}{g'} \right]\right)\right]\nn\\
&+ \frac{k}{4\pi} \int dt d\vp \, \left[ \frac{2}{3}\bigg((k_0-k_1)^2 f'\partial_-f + (k_0+2k_1)^2 g'\partial_-g \right. \nn\\
&~~~~~~~~~~~~~~~~~~~~~~~~+ \left.\frac{(k_0-k_1)(k_0+2k_1)}{2}\left[f'\partial_-g +g'\partial_-f\right]\bigg)\right]\nn\\
&+\frac{k}{4\pi} \int dt d\vp \, \left[ -\frac{2}{3}\left[(k_0-k_1) f'\frac{\partial_- g'}{g'}+(k_0+2k_1) g'\frac{\partial_- f'}{f'}\right]\right]\,.
\end{align}

This is the action on the outer boundary of an annulus (trumpet) geometry in the presence of dynamical holonomies (zero modes). We claim that this action is the action on the symplectic leaves of the $W_3$ Poisson manifold for the constant representatives $k_0$ and $k_1$. It generalizes the result of \cite{Marshakov:1989ca} in the presence of non-trivial holonomies. Indeed when there is no non-trivial holonomy, i.e. $k_0=k_1=0$, the Hamiltonian reduces to the two-dimensional generalization of the $W_3$ Schwarzian action \cite{Marshakov:1989ca}.

%Moreover, in the case that either $k_0$ or $k_1$ vanishes, the last two terms in the action \eqref{eq:action_schw-W3} will be just a total derivative. This explains why these terms do not appear in the case of $SL(2)$ Chern-Simons theory.    

At this point lets point out a few interesting observations about these boundary actions. The first remark is regarding the field redefinitions \eqref{eq:field_redef_W3_1} and \eqref{eq:field_redef_W3_2}. Since $\gamma_1$ and $\gamma_2$ are periodic, one can immediately notice that $f(t,\vp)$ and $g(t,\vp)$ get a $2\pi$ shift when going around a circle, i.e.
\begin{align}
f(t,\vp+2\pi) &= f(t,\vp) +2\pi,\nonumber\\
g(t,\vp+2\pi) &= g(t,\vp) +2\pi.
\end{align}
This is indeed very similar to $SL(2)$ Chern-Simons theory where such a field appears in the Schwarzian action. In that case $f(t,\vp)$ was an element of reparametrization of the circle or equivalently an element of ${\rm Diff}(S^1)$. However, for $SL(3)$ theory the set $(f,g)$ is an element of diffeomorphisms of $\rm{RP}^2$ real projective plane \cite{Ovsienko1990}. In fact, for $SL(n)$ theory the Schwarzian action is written in terms of fields parametrizing symplectic leaves of ${\rm Diff}(\rm{RP}^{n-1})$. Notice that $\rm{RP}^{1} \cong S^1$.

Another remark is regarding the invariance of the $SL(3)$ Schwarzian action. In the case of $SL(2)$, there was a $U(1)$ gauge redundancy which is the isotropy group of $SL(2)$ for the hyperbolic conjugacy class. This resulted in the invariance of the $SL(2)$ Schwarzian action under ${\rm Diff}(S^1)/S^1$. For more details, see \cite{Henneaux:2019sjx} and section \ref{sec:geometric-description} for the same discussion from the perspective of the geometric action.  The isotropy group of $SL(3)$ for the non-degenerate hyperbolic holonomy is $\rm{R}^* \times \rm{R}^*$ where $\rm{R}^*$ is the projectively extended real line (one point compactification of $\rm{R}$). It is topologically isomorphic to $S^1$. Another way to realize this isotropy group is the gauge redundancy \eqref{eq:gauge_red_W3} which is $U(1) \times U(1)$. Therefore, the Schwarzian action \eqref{eq:action_schw-W3} is invariant under ${\rm Diff}(\rm{RP}^2)/\rm{R}^* \times \rm{R}^* \cong {\rm Diff}(\rm{RP}^2)/S^1 \times S^1$ .

%COMMENTS:

%1) I am highly critical of these extra terms as I do not have any geometrical understanding of them.

%2) The action on the coadjoint orbit is usually written in terms of a Schwarzian derivative and the extra term which involves the orbit representative. The Schwarzian derivative, however, appears in the transformation of Energy-Momentum tensor under finite local conformal symmetries. The Energy-Momentum tensor ($\mathcal{L}$) is written in terms of Virasoro modes, which is the highest quasi-primary modes in the conformal field theory. If we consider presence of higher spin modes which is the case in $W$ algebras, one should presumably study how higher modes transform under finite local conformal symmetries. They should not contribute to the Hamiltonian. Do they appear somehow in the action?

\subsubsection{Diagonal Embedding} 
We can also choose a basis for the $\mathfrak{sl}(3)$ representation known as the diagonal embedding, in which the $\mathfrak{sl}(3)$ generators satisfy the algebra \cite{Bunster:2014mua}  
\begin{align}\label{eq:diagonal_embed}
[U_m, U_n] &= (m-n) U_{m+n},\\
[U_m, J_0] &=0,\\
[U_m, G_n^{\pm}] &= (\frac{m}{2}-n) G_{m+n}^{\pm},\\
[J_0,G_m^{\pm}]& = {\pm}\, G_m^{\pm},\\
[G_m^+,G_n^-] &= U_{m+n} -\frac{3}{2}(m-n) J_0.
\end{align}

A matrix representation for these generators is 
\begin{align}
U_{-1} &= \begin{pmatrix} 0 & 0 & -1 \\ 0 & 0 & 0 \\ 0 & 0 & 0 \end{pmatrix}, \qquad~ U_{0} = \frac{1}{2}\begin{pmatrix} 1 & 0 & 0 \\ 0 & 0 & 0 \\ 0 & 0 & -1 \end{pmatrix}, \qquad U_{+1} = \begin{pmatrix} 0 & 0 & 0 \\ 0 & 0 & 0 \\ 1 & 0 & 0 \end{pmatrix},\nn\\
J_0 &= \begin{pmatrix} \frac{1}{3} & 0 & 0 \\ 0 & -\frac{2}{3} & 0 \\ 0 & 0 & \frac{1}{3} \end{pmatrix}, \qquad G_{+1/2}^+ = \begin{pmatrix} 0 & 0 & 0 \\ 0 & 0 & 0 \\ 0 & 1 & 0 \end{pmatrix}, \qquad G_{+1/2}^- = \begin{pmatrix} 0 & 0 & 0 \\ 1 & 0 & 0 \\ 0 & 0 & 0 \end{pmatrix},\nn\\
G_{-1/2}^+ &= \begin{pmatrix} 0 & 1 & 0 \\ 0 & 0 & 0 \\ 0 & 0 & 0 \end{pmatrix}, \qquad G_{-1/2}^- = \begin{pmatrix} 0 & 0 & 0 \\ 0 & 0 & -1 \\ 0 & 0 & 0 \end{pmatrix}.\nn
\end{align}

The highest and lowest weight gauge conditions in terms of the $\mathfrak{sl}(2)$-diagonal embedding basis take the following form
\begin{align}
\cA_r &= 0,\, \cA_\vp = U_{+1} - \frac{8\pi}{k}\left[ \left(\cL(t,\vp)-\frac{6\pi}{k}\mathcal{U}(t,\vp)\right)U_{-1} + \frac{3}{2}\,\mathcal{U}(t,\vp) J_0 + \Psi_{\pm } G^{\pm}_{-1/2} \right]\, \textrm{on}\,\Sigma_o\nn\\
\cA_r &= 0,\,\cA_\vp = U_{-1} - \frac{8\pi}{k}\left[ \left(\cL(t,\vp)-\frac{6\pi}{k}\mathcal{U}(t,\vp)\right)U_{+1} + \frac{3}{2}\,\mathcal{U}(t,\vp) J_0 + \Psi_{\pm } G^{\pm}_{+1/2}\right]\,\textrm{on}\,\Sigma_i.\nn
\end{align}
At the outer boundary, this results in the following conditions
\begin{align}
-\frac{1}{2} e^{\phi_1}F(\gamma_1,\gamma_2,\gamma_3;k_0,k_1)&= 1,\\
e^{\frac{\phi_1 +3 \phi_2}{2}} \left[\gamma_1' - (k_0 -k_1)\gamma_1\right]-\xi_2 &=0,\\
 e^{\frac{\phi_1 - 3 \phi_2}{2}} \left[\gamma_2' - (k_0+2k_1)\gamma_2\right] + \xi_1 &=0.
\end{align}

Inserting these conditions into the action \eqref{eq:outer_boundary}, we obtain the reduced action on the outer boundary
\begin{align}
S_{\rm bdy}^{\Sigma_o}[k_0, k_1,\phi_1,\phi_2] =  \frac{k}{4\pi} \int dt d\vp \; &\bigg(\frac{1}{2}\partial_- \phi_1 (\phi'_1 + 4 k_0 + 2 k_1) + \frac{3}{2}\partial_- \phi_2 (\phi'_2 -2 k_1)\nn\\
&+ \left(\xi_2 \partial_- \xi_1 -\xi_1 \partial_- \xi_2\right)- 2(k_0^2 + k_0 k_1 + k_1^2) \bigg).
\end{align}
Following the same steps at the inner boundary, the complete boundary action can be written as
\begin{align}\label{eq:final_action_W3_diag}
S_{\rm bdy} =  \frac{k}{4\pi} \int dt d\vp \, &\bigg(\frac{1}{2}\partial_- \phi_1 \,\phi'_1 - \frac{1}{2}\partial_+ \tilde{\phi}_1 \, \tilde{\phi}'_1 
 +\frac{3}{2}\partial_- \phi_2 \, \phi'_2 - \frac{3}{2}\partial_+ \tilde{\phi}_2 \, \tilde{\phi}'_2 \nn\\
&+ \left(\xi_2 \partial_- \xi_1 -\xi_1 \partial_- \xi_2\right) - \left(\tilde{\xi}_2 \partial_+ \tilde{\xi}_1 -\tilde{\xi}_1 \partial_+ \tilde{\xi}_2\right)\nn\\
& + (2k_0 + k_1)(\partial_- \phi_1- \partial_+ \tilde{\phi}_1) - 3 k_1(\partial_- \phi_2 - \partial_+ \tilde{\phi}_2)\nn\\ 
&- 4(k_0^2 + k_0 k_1 + k_1^2) \bigg),
\end{align}
where $\tilde{\phi}_1,\tilde{\phi}_2,\tilde{\xi}_1,\tilde{\xi}_2$ are the fields on the inner boundary.

It is interesting to notice that the second line is the free field theory of spinors with bosonic statistics. It can be understood by comparing this action to the one of $\mathcal{N}=2$ supergravity \eqref{eq:S_complete_any_N} and setting the $R$-symmetry to zero. The sign difference between these two descriptions is because the fields $\xi_1,\xi_2,\tilde{\xi}_1,\tilde{\xi}_2$ are not Grassmann variables, unlike the corresponding fields in supergravity.

The action \eqref{eq:final_action_W3_diag} has a gauge redundancy given by
\begin{align}\label{eq:gauge_red_W3_diag}
\phi_1 &\rightarrow \phi_1 + \epsilon_1(t), \qquad \tilde{\phi}_1 \rightarrow \tilde{\phi}_1 + \epsilon_1(t)\,,\nonumber\\
\phi_2 &\rightarrow \phi_2 + \epsilon_2(t), \qquad \tilde{\phi}_2 \rightarrow \tilde{\phi}_2 + \epsilon_2(t)\,,\nonumber\\
\xi_{1,2} &\rightarrow \xi_{1,2},\qquad \qquad ~ \tilde{\xi}_{1,2} \rightarrow \tilde{\xi}_{1,2} \,,\nonumber\\
k_0 &\rightarrow k_0,\qquad \qquad ~~~~ k_1 \rightarrow k_1 \,.
\end{align}

This is once again related to the fact that the isotropy group of $SL(3)$ group for the non-degenerate holonomy conjugacy class is given by $\rm{R}^* \times \rm{R}^* \cong U(1) \times U(1)$. Therefore the action \eqref{eq:final_action_W3_diag} is invariant under $\rm{Diff}(\rm{RP}^2)/S^1 \times S^1$. The Schwarzian can be obtained by an appropriate change of variables. However, we will not present it here and we end this discussion by pointing out that given the similarities between the action \eqref{eq:final_action_W3_diag} and the action for $\cN =2$ Neveu-Schwarz supergarvity \eqref{eq:S_complete_any_N} one can find the change of variables by following the discussion of $\cN =2$ supergravity in section \ref{sec:N=2}.

It is important to make a comment about the boundary action \eqref{eq:final_action_W3_diag}. This action which is obtained through the Drinfeld reduction should be thought as the candidate for the action on the symplectic leaves of $W_3$ manifold. Even though it is straightforward to write it down from the Hamiltonian reduction, it is not obvious how to obtain it in the geometric approach. And moreover we consider the presence of non-trivial holonomies which not only allows for solutions like BTZ but also provides a geometric sense. This is because the holonomies are related to the representatives of symplectic leaves and the invariant charges of the theory, $\cL(t,\vp), \mathcal{U}(t,\vp), \Psi_{\pm}(t,\vp)$, are determined by the intersections of the symplectic leaves.  

In the next section, we show explicitly the construction of Schwarzian boundary actions for $\mathcal{N}=1,2$ supergravity with the Kirillov-Kostant construction of geometric actions. Given that for these theories the symplectic leaves are the coadjoint orbits, the procedure is tedious but straightforward. We will show that these geometric actions match the ones obtained through Hamiltonian reduction where non-trivial holonomies appear as the orbit representatives.

%%%%%%%%%%%%%%%%%%%%%%%%%%%%%%%%%%%%%%%%%%%%%%%%%%%%%%%%%%%%%%%%%%%%%%%%%%%%%%%%%%%%%%%%%%%%%%%%%%%%%%%%%%%%%%%%%%%%%%%%%%%%%%%%%%%%%%%%%%%%%%%%%%%%%%
\section{Diffeomorphisms of the supercircle}\label{sec:geometric-description}

In this section we study explicitly the case of $\cN = 0, 1$ and $2$ from the general results of sections \ref{sec:N1superCS} and \ref{sec:N2superCS} and we will show that the actions \eqref{S_reduced_full} and \eqref{eq:S_complete_any_N} can be understood as the geometric action on the coadjoint orbit of the group of reparameterizations of the supercircle. The bosonic case proceeds similarly to the case discussed in \cite{Barnich:2017jgw,Cotler:2018zff}, although here we clarify the role of the bulk holonomy as the orbit representative. The $\cN=1$ supersymmetric case was also discussed in \cite{Cotler:2018zff}. Here we generalize the results to arbitrary bulk holonomy and extend it to $\cN =2$ supersymmetry. 

The case of $\cN > 2$ is interesting because the asymptotic symmetry algebra \eqref{superconforml_alg} becomes non-linear. Due to this non-linearity there is no clear interpretation in terms of coadjoint orbit of some supergroup. Instead, one can think about these cases by suitably generalizing the concept of symplectic leaves of a Poisson manifold, which in the case of linear algebras corresponds to the coadjoint orbits. We will briefly comment on this in Appendix \ref{app:sym_leaves}. In this section, we start by reviewing the Kirillov-Kostant construction of geometric actions on the coadjoint orbits in the case of the centrally extended groups of reparameterizations of the supercircle.

\subsection{Reparameterizations of the (super)circle}

%The Alekseev-Shatashvili action is the geometric action on the coadjoint orbit of the centrally extended group of diffeomorphisms of the circle; the Virasoro group. We hence can interpret the boundary action as one for reparameterizations of the boundary circle. Since we took the holonomy of the Chern-Simons connection fixed, which means the gravitational solution has fixed mass and angular momentum, this theory describes the Brown-Henneaux diffeomorphisms around a given bulk solution. The corresponding theory is equivalent to an action on the coadjoint orbit of the Virasoro group, with the orbit representative given by the holonomies of the Chern-Simons connection.

The coadjoint action on a generic element of the dual space $b_0 \in \mathfrak{g}^*$ of any Lie algebra defines a particular coadjoint orbit. The coadjoint orbits are symplectic manifolds isomorphic to the coset $G/\cH_{b_0}$, where $G$ is the Lie group in question and $\cH_{b_0}$ the stabilizer subgroup of the orbit, defined by all elements $h \in G$ which leave the orbit representative $b_0$ invariant under the coadjoint action ${\rm Ad}_{h}^*$. The symplectic form on the orbit is the Kirillov-Kostant symplectic form $\Omega$. As it is closed, locally it is also exact $\Omega = \extd \alpha$. Then $\alpha $ can be used to define an action on the orbit $ S[g;b_0] = \int_{\gamma} \alpha$, whose symplectic structure by definition coincides with the Kirillov-Kostant symplectic structure. This action is known as the geometric action on the coadjoint orbit. The addition of invariant Hamiltonians was discussed in \cite{Barnich:2017jgw} and the general construction is reviewed in appendix \ref{sec:geomaction}. From there we see that the Kirillov-Kostant symplectic form for centrally extended groups consists of two terms; one proportional to the orbit representative $b_0$ and one proportional to the central charge $c$
\begin{align}\label{Omega}
\Omega = \extd \alpha &  =  \extd  \langle \Ad^*_{g^{-1}} b_0 , Y \rangle - \frac{c}{2} \langle \extd S(g), Y \rangle \, \\
& =   \extd  \langle \Ad^*_{g^{-1}} (b_0,c), (Y, n_Y) \rangle .\nonumber
\end{align}
Here $\Ad^*_{g^{-1}} b_0$ is the coadjoint action of $g \in G$ on the orbit representative $b_0$, $\langle , \rangle$ is the pairing between the Lie algebra and its dual space and hence defines a map $\mathfrak{g}^* \times \mathfrak{g} \to \mathbb{R}$. $S(g)$ is the Souriau cocycle defining the central extension of $G$ to $\hat{G} = G \times \mathbb{R}$.

In the above equation, $n_Y$ solves equation \eqref{nYeqn} and $Y$ can be obtained from the following equality
\begin{equation}\label{Ydef}
\extd b = - \ad_Y^* b.
\end{equation}
For $b = \Ad_{g^{-1}}^* b_0$ this gives
\begin{equation}\label{Yeqn3}
\extd (\Ad_{g^{-1}}^* b_0)  = - \ad_Y^* (\Ad_{g^{-1}}^* b_0).
\end{equation}
From \eqref{Ydef} it follows that $Y$ solves the Maurer-Cartan equation $\extd Y = - \frac12 \ad_Y Y$, whose solution is locally $Y = g^{-1}dg$. So to find the geometric action, one needs to compute the Maurer-Cartan form $Y$ and write the Kirillov-Kostant symplectic form \eqref{Omega} as a total exterior derivative, or find $n_Y$ as a solution to \eqref{nYeqn}. In practice it will also be useful to do the former and to obtain $Y$ from \eqref{Yeqn3}.

The Kirillov-Kostant symplectic form will give the symplectic part of the action. The geometric action can be extended by adding the Noether charge for a global symmetry as Hamiltonian \cite{Barnich:2017jgw}. The resulting action is by construction invariant under the same symmetries as the symplectic term. The Noether charge associated to a symmetry generated by a vector field $(X,n)$ in the centrally extended Lie algebra $\hat{\mathfrak{g}} = \mathfrak{g} \oplus \mathbb{R}$ is given by
\begin{equation}
Q_{(X,n)} = - \langle (b,c), (X,n) \rangle\,.
\end{equation}
Hence as Hamiltonian we may simply add
\begin{equation}\label{Hamiltonian_KK}
H_{(X,n)} = - \int_\gamma\; Q_{(X,n)},
\end{equation}
for $\gamma$ a path along the orbit. The total geometric action as a function of group elements $g$ and for a given $(b_0,c)$ is then  \cite{Aratyn:1989qq}
\begin{equation}\label{geomaction}
S_{\hat{G}}[g;b_0,c] = \int_{\gamma} \alpha -  Q_{(X,n)} = \int_{\gamma} \langle \Ad_{g^{-1}}^* (b_0,c), (Y,n_Y) - (X,n) \rangle\,.
\end{equation}
We will now work out the different ingredients in this expression for the centrally extended group of diffeomorphisms of the supercircle.

\subsubsection{Geometric action for superconformal groups}
When the Lie group is taken to be the centrally extended group of diffeomorphisms of the supercircle $\widehat{\rm Diff} (S^{1|\cN})$, the geometric action constructed from \eqref{Omega} corresponds to the kinetic term of the reduced AdS$_3$ supergravity action \eqref{eq:S_complete_any_N} up to the $\cN =2$ case, as we will now demonstrate.
Let us denote the general $\cN$-extended superspace by superspace coordinate $z = \{\vp, \theta^i\}$ where $0 < \vp \leq 2\pi$ is a bosonic coordinate (the angle on the circle) and $\theta^i$ is a collection of $\cN$ Grassmann coordinates. The supercovariant derivative is
\begin{equation}
D^i = \partial_{\theta^i} + \theta^i \partial_\vp\,,
\end{equation}
such that
\begin{equation}
\{ D^i ,D^j\} = 2 \delta^{ij} \partial_\vp\,.
\end{equation}
Reparameterizations of the supercircle consists of coordinates transformations
\begin{equation}
\vp \to \tilde \vp (\vp,\theta^i) \,, \qquad \theta^j \to \tilde \theta^j (\vp, \theta^i)\,,
\end{equation}
subject to the constraint that the supercovariant derivative transforms covariantly, or
\begin{equation}\label{Dtrafo}
D^i = (D^i \tilde \theta^j) \tilde D^j\,.
\end{equation}
This condition implies the following useful identities
\begin{align}
D^i \tilde \vp & = \tilde \theta D^i \tilde \theta ,\label{id1} \\
(D^i \tilde{\theta}^j) (D^k \tilde \theta^j) & = \delta^{ik} ( \partial_\vp \tilde \vp + \tilde \theta^j \partial_\vp \tilde \theta^j), \label{id2} \\
\det \left(D \tilde \theta\right) & = \pm ( \partial_\vp \tilde \vp + \tilde \theta^j \partial_\vp \tilde \theta^j)^{\cN/2} .\label{id3}
\end{align}
Elements of the super-Virasoro algebra consists of the pair $(X(z), n)$ and this is paired with elements of the dual space $(B(z),c)$ (we will denote the superspace dual elements by capital $B$, keeping $b$ and $b_0$ reserved for the bosonic orbit representative) as
\begin{equation}
\langle (B,c), (X,n) \rangle = \langle b, X \rangle + cn = \int \extd z \; B(z) X(z) + cn\,,
\end{equation}
where $\extd z = \extd \vp \extd \theta^1 \ldots \extd\theta^\cN$.
The adjoint and coadjoint action of the centrally extended superconformal group are
\begin{align}
\Ad_{\tilde z} (X(z),n) & = \left( \det(D\tilde \theta)^{-\frac{2}{\cN}} X(\tilde z) , n - \int \extd z \, S^{(\cN)}(z;\tilde z^{-1}) X(z) \right) \,,\\
\Ad_{\tilde z}^* (B(z), c) & = \left( \det(D\tilde \theta)^{\frac{4-\cN}{\cN}} B(\tilde z) - c S^{(\cN)}(z;\tilde{z}) ,c \right) \,. \label{coad_action}
\end{align}
Here $S^{(\cN)}(z;\tilde{z})$ are the super Schwarzian derivatives for $\cN$ extended supersymmetry. They satisfy the cocycle condition
\begin{equation}\label{supercocycle}
S^{(\cN)}(z,\tilde{\tilde{z}}) = (\det(D\tilde \theta))^{\frac{4-\cN}{\cN}} S^{(\cN)}(\tilde z, \tilde{\tilde{z}}) + S^{(\cN)}(z,\tilde{z})
\end{equation}
and $S^{(\cN)}(z;z) = 0$ so that $S^{(\cN)}(z;\tilde z) = - \Ad_{\tilde{z}}^* S^{(\cN)}(z;\tilde{z}^{-1})$. The condition \eqref{supercocycle} can be solved only for $\cN \leq 4$ \cite{Schoutens:1988ig} and the limiting $\cN=4$ case is special as there are two independent non-trivial (and non-local) solutions. The upper bound $\cN=4$ is natural as the stress tensor sits in a supermultiplet of conformal dimension $(2- \frac{\cN}{2})$, which becomes negative for $\cN>4$. In our case, however, the asymptotic symmetry algebras \eqref{superconforml_alg} correspond to the superconformal Lie algebras of $\widehat{\rm Diff}(S^{1|\cN})$ only for $\cN = 0,1,2$. For $\cN > 2$ there is a non-linear term in the superconformal algebras \eqref{superconforml_alg}, which makes the interpretation in terms of coadjoint orbits difficult. Hence here we will discuss the case up to $\cN =2$.

The result for the (super)-Schwarzians is
\begin{align}
& \cN=0 & S^{(0)}(\vp; \tilde \vp) & = \frac{1}{24\pi} \left( \frac{\tilde \vp'''}{\tilde \vp'} - \left(\frac{\tilde \vp''}{\tilde \vp'}\right)^2 \right) \,, \label{N0Schwarz} \\
& \cN=1 & S^{(1)}(z; \tilde z) & =  \frac{1}{12\pi} \left( \frac{\partial_\vp^2 \tilde \theta }{D\tilde \theta} - 2 \frac{D \partial_\vp\tilde \theta \partial_\vp \tilde \theta}{(D\tilde \theta)^2} \right)\,,  \label{N1Schwarz} \\
& \cN=2 & S^{(2)}(z; \tilde z) & =  \frac{1}{24\pi} \epsilon^{ij}  \frac{2( D^i \partial_\vp \tilde \theta^k)( D^j \tilde \theta^k) - 2 ( D^i D^j \tilde \theta^k)  (\partial_\vp \tilde \theta^k) }{(D^p\tilde \theta^q) (D^p\tilde \theta^q)}\,.   \label{N2Schwarz}
% \\
%& N=3 & S^{(3)}(z; \tilde z) & =   \frac{1}{12\pi} \epsilon^{ijk}  \frac{( D^i D^j \tilde \theta^l)( D^k \tilde \theta^l) }{(D^p\tilde \theta^q) (D^p\tilde \theta^q)} \\
%& N=4\;  {\rm (regular)} & D^i S^{(4)}(z; \tilde z) & =  \frac{1}{9\pi} \epsilon^{ijkl}  \frac{( D^j D^k \tilde \theta^m)( D^l \tilde \theta^m) }{(D^p\tilde \theta^q) (D^p\tilde \theta^q)} \\
%& N=4\; {\rm (twisted)} & D^i S^{(4)}(z; \tilde z) & =  \frac{2}{\pi} \frac{( \partial_x \tilde \theta^j)( D^i \tilde \theta^j) }{(D^p\tilde \theta^q) (D^p\tilde \theta^q)} - \frac{1}{3\pi} \frac{( D^i D^j \tilde \theta^k)( D^j \tilde \theta^k) }{(D^p\tilde \theta^q) (D^p\tilde \theta^q)}
\end{align}
To proceed we need to find the Maurer-Cartan form. We can do so by solving for the $\partial_{\tilde \vp} B_0$ part of \eqref{Yeqn3}. For the centrally extended group of diffeomorphisms of the supercircle, the relevant part of \eqref{Yeqn3} becomes
\begin{equation}
\extd \left( \det(D\tilde \theta)^{\frac{4-\cN}{\cN}} B_0(\tilde z) - c S^{(\cN)}(z;\tilde{z}) \right) = - \ad_Y^* \left( \det(D\tilde \theta)^{\frac{4-\cN}{\cN}} B_0(\tilde z) - c S^{(\cN)}(z;\tilde{z}) \right).
\end{equation}
The $B_0, \partial_{\tilde \vp}B_0, \tilde D^i B_0$ and $c$ dependent parts of this equation should vanish individually. Expanding this equation using the following useful formulas
\begin{align}
\ad_Y X &= Y \partial_\vp X - X \partial_\vp Y + \frac12 D^i Y D^i X,\\
\extd B_0(\tilde \vp, \tilde \theta^i) &= \partial_{\tilde \vp} B_0(\tilde \vp, \tilde \theta^i) (\extd \tilde \vp + \tilde \theta^j \extd \tilde \theta^j ) + \extd \tilde  \theta^j \tilde D^j B_0(\tilde \vp, \tilde \theta^i),
\end{align}
and 
\begin{equation}
\partial_\vp = \partial_\vp \tilde \theta^k \tilde D^k + (\det D\tilde \theta)^{2/\cN} \partial_{\tilde \vp},
\end{equation}
we find an over-constrained set of equations for $Y$ which is solved by
\begin{equation}
Y = - \frac{\extd \tilde \vp + \tilde \theta^i \extd \tilde{\theta}^i}{(\det D\tilde{\theta})^{2/\cN}}.
\end{equation}
We now have all the ingredients to write down the geometric actions on the coadjoint orbit for the $\cN$-extended superconformal group. The $c$-independent part of the symplectic term can be written down for generic $\cN$. It is (writing $\extd = \extd t \, \partial_t$):
\begin{equation}\label{Igeomb0}
 - \int \extd t \; \extd z \; \det (D\tilde \theta)^{\frac{2-\cN}{\cN}} B_0(\tilde \vp, \tilde{\theta}^i) (\dot{\tilde{\vp}} + \tilde \theta^j \dot{\tilde{\theta}}^j )
\end{equation}
The $c$-dependent part should be obtained by writing the second term in \eqref{Omega} as a total exterior derivative for the relevant Schwarzian listed in \eqref{N0Schwarz}-\eqref{N2Schwarz}. We will study this on a case by case basis in the next sections. 

Before doing so we comment on the Hamiltonian. In the case of interest the action is invariant under a shift symmetry generated by vector fields $ - \partial_\vp$. So we may add to the symplectic part a Hamiltonian \eqref{Hamiltonian_KK} with $(X,n) = (-1,0)$.\footnote{The symmetry generator $(X,n)= (0,n)$ generate constant shifts of $n$ which are trivial as the vectors belong to the extended little algebra \cite{Barnich:2017jgw}.} This, together with \eqref{coad_action} gives the Hamiltonians
\begin{equation}\label{HamgenN}
H^{(\cN)}(\tilde{z}; B_0,c) = - \int \extd t \, \extd z \left( \det (D\tilde \theta )^{\frac{4-\cN}{\cN}} B_0 (\tilde z) - c S^{(\cN)}(z; \tilde z) \right).
\end{equation}
Due to the cyclic property of the Schwarzian derivative \eqref{supercocycle}, the orbit representative term of the Hamiltonian \eqref{HamgenN} can be absorbed into the super-Schwarzian by a second reparameterization of the supercircle $\tilde{\tilde{z}}$ such that
\begin{equation}\label{orbitredef}
c S^{(\cN)}(\tilde{z}; \tilde{\tilde{z}}) = - B_0 (\tilde{z})\,.
\end{equation}
The Hamiltonian then becomes $H^{(\cN)}(\tilde{\tilde{z}};0,c) =  \int \extd t \, \extd z \left(  c S^{(\cN)}(z; \tilde{\tilde{z}}) \right)$ and the orbit representative terms will translate into periodicity conditions of the new variable $\tilde{\tilde{z}}$. The same transformation will remove the orbit representative term \eqref{Igeomb0} from the symplectic term \cite{Barnich:2017jgw}.

Next we will discuss explicitly the $\cN=0, 1,2$ cases to show that the actions \eqref{S_reduced_full} and \eqref{eq:S_complete_any_N} are indeed the geometric actions on the coadjoint orbit of the centrally extended group of diffeomorphisms of the supercircle. The orbit representative terms are given by the holonomy of the Chern-Simons connection by comparing the periodicities of the fields to those of $\tilde{\tilde{z}}$ satisfying \eqref{orbitredef}.

%%%%%%%%%%%%%%%%%%%%%%%%%%%%%%%
%%%%%%%%%%%%%%%%%%%%%%%%%%%%%%%

\subsection{Bosonic case}\label{sec:bosonic_WZW}

The geometric action on the coadjoint orbit of the Virasoro group of diffeomorphisms of the circle $\widehat{\rm Diff}(S^1)$ was worked out by Alekseev and Shatashvili \cite{Alekseev:1988ce} and reported to be relevant to three dimensional AdS$_3$ gravity in \cite{Barnich:2017jgw} and later expanded upon in \cite{Cotler:2018zff}. In \cite{Henneaux:2019sjx} it was shown that for the bosonic theory, the action obtained from the Hamiltonian reduction coincides with Alekseev-Shatashvili action where the holonomy plays the role of orbit representative. In order to set the ground for the supersymmetric case, we briefly review the discussion here. 

From the last two subsections, we reproduce the Alekseev-Shatashvili action from \eqref{geomaction} in the bosonic case. This is
\begin{equation}\label{ASaction}
S_{\widehat{\rm Diff}(S^1)}[f(\vp);b_0,c] =   \int dt d\vp \; \left[\frac{c}{24\pi} \left( \frac32 \frac{f'' \partial_- f'}{f'^2}-\frac{\partial_- f''}{f'} \right) + b_0(f) f' \partial_- f \right],
\end{equation}
where we have taken $\tilde{\vp} = f(\vp)$ subject to the periodicity condition $f(\vp+2 \pi) = f(\vp) + 2\pi$.
From \eqref{S_reducedn=2} we see that in the bosonic case (i.e., no fermions, no $R$-symmetry and $k_r=0$) the reduction from Chern-Simons theory on the outer boundary gives
\begin{equation}
S =  \frac{k}{8\pi} \int  \extd t \,\extd \vp \; \bigg[ \Big(\partial_- \log(Y'-k_0 Y) \Big)\Big( \partial_\varphi \log(Y'-k_0 Y)\Big) - k_0^2\bigg]\,,
\end{equation}
up to total derivative terms. Moreover, in order to make a link with the classification of (super-)Virasoro coadjoint orbits, in this section we focus on the case where the holonomy is constant and time-independent, which is the case for constant orbit representatives.
However, the field $Y(\vp)$ here is periodic, i.e. $Y(\vp+2\pi)=Y(\vp)$ which is not the usual $Y(\vp + 2\pi) = Y(\vp) + 2\pi$ as it is for elements of $\widehat{\rm Diff}(S^1)$. Therefore, in order to relate these two actions we consider a field transformation
\begin{equation}
	\tilde{\tilde{\vp}} \equiv Y(f(\vp)) = e^{-k_0 (f(t,\vp)-\vp)}.
\end{equation}
such that $f(\vp + 2\pi) = f(\vp) + 2\pi$ while $Y(\vp)$ is periodic. In the new variable $f(\vp)$ the action takes the form of \eqref{ASaction} with the central charge and the orbit representative given as
\begin{equation}
    c= 6k,\qquad\qquad\qquad b_0 = \frac{k}{8\pi} k_0^2.
\end{equation}
In the case in hand, the holonomy is in the hyperbolic conjugacy class and therefore we have $k_0^2 = 4\cL_0$. One then finds that the orbit representative $b_0$ is related to the zero mode of the bulk Chern-Simons solution as
\begin{equation}\label{b0L0}
b_0 = \frac{c}{12\pi} \cL_0 = \frac{k}{2\pi} \cL_0\,.
\end{equation}
We have thus illustrated that the zero mode of the bulk solutions correspond directly to the constant representatives of the Virasoro coadjoint orbits. In the case of the BTZ solutions, where $\cL_0$ is positive, these correspond to the generic Virasoro orbit with a $U(1)$ little group generated by the action of $L_0$. The Noether charge for this $L_0$ action is exactly the Hamiltonian \eqref{HamgenN} that was added to the geometric action. 

The exceptional orbits of Virasoro have constant representative terms $b_0 = - \frac{c n^2}{48 \pi}$ for integer $n$. At these points, the little group on the orbit is enlarged to $SL^{(n)}(2,\mathbb{R})$. From the bulk perspective these orbits correspond to (the left-moving sector of) Ba\~nados geometries with $\cL_0 = - \frac14 n^2$ for integer $n$. 
For $n=1$ this is the global AdS$_3$ ground state, while for larger $n$ the geometries have an angular access of a multiple of $2\pi$. For those solutions the Hamiltonian of \eqref{ASaction} becomes unbounded from below \cite{Witten:1987ty} and hence this is a good reason to discard them.

The bulk solutions with conical defects correspond to orbits with negative $b_0 \neq - \frac{cn^2}{48\pi}$. At this point $k_0 = \sqrt{4\cL_0}$ becomes imaginary and hence the holonomy $h = \exp(4\pi \sqrt{\cL_0} L_0)$ becomes complex and is not an element of $SL(2,\mathbb{R})$ anymore. This is the Chern-Simons perspective for discarding solutions with conical singularities. For the exceptional orbits where $b_0 = - \frac{c n^2}{48 \pi}$ the holonomy is still an element of $SL(2,\mathbb{R})$ since exactly at those values we have $h = \exp(2\pi n i L_0)$. The complete discussion of the geometric action with holonomies in different conjugacy classes of $SL(2,\mathbb{R})$ can be found in \cite{Henneaux:2019sjx}.

\subsection{$\cN=1$ supersymmetry}\label{sec:N=1}

Let us now consider the centrally extended group of diffeomorphisms of the $\cN=1$ supercircle $\widehat{\rm Diff}(S^{1|1})$. Writing the geometric action \eqref{geomaction} using the $\cN=1$ super-Schwarzian \eqref{N1Schwarz} we find the superspace action worked out in \cite{Aratyn:1989qq}, up to total derivative terms
\begin{align}\label{N1superspaceaction}
S_{\widehat{\rm Diff}(S^{1|1})} (\tilde{z}; B_0,c) %& = \int \extd t\, \extd z \bigg[ \frac{c}{6\pi} \left( \frac{\partial_\vp \partial_+\tilde \theta}{D\tilde{\theta}} - 2 \frac{\partial_\vp D \tilde \theta \, \partial_+ \tilde \theta}{(D\tilde \theta)^2}\right)- B_0(\tilde z) D\tilde{\theta} (\partial_+ \tilde \vp + \tilde{\theta} \partial_+ \tilde \theta) \bigg] \\
& =  \int \extd t\, \extd z \bigg[ \frac{c}{12\pi} \left( \frac{\partial_\vp D \tilde \theta \, \partial_- \tilde \theta}{(D\tilde \theta)^2}\right) + B_0(\tilde z) D\tilde{\theta} (\partial_- \tilde \vp + \tilde{\theta} \partial_- \tilde \theta) \bigg] 
\end{align}
In order to compare this to the action obtained from the Chern-Simons reduction \eqref{S_reducedn=2}, we need a suitable parameterization of $\tilde \theta $ to perform the superspace integral and write the $\cN=1$ superspace action in components.

The $\cN=1$ constraint $D \tilde{\vp} = \tilde{\theta} D \tilde{\theta}$ is solved by parameterizing the diffeomorphism of the supercircle as
\begin{align}\label{N=1repara}
\tilde{\vp} & = f(\vp + \theta \eta(\vp)) \\
\tilde{\theta} & = \sqrt{f'(\vp)} \left( \eta(\vp) + \theta \left(1 + \frac12 \eta \eta' \right) \right)
\end{align}
Here $f(\vp + 2\pi) = f(\vp) + 2 \pi$. 
Taking the orbit representative term to be purely bosonic, i.e. $B_0(\tilde{z}) = b_0(f) \tilde{\theta}$ and using the above parameterization, the action \eqref{N1superspaceaction} becomes (up to total derivatives)
\begin{align}\label{N1supervirgeom}
S_{\widehat{\rm Diff}(S^{1|1})}(f,\eta;b_0,c)  =  \int dt d\vp & \; \Bigg[ -\frac{c}{24\pi}\bigg(   \frac{\partial_-{f}''}{f'} - \frac32 \frac{\partial_- f' f''}{f'{}^2} - S^{(0)}(\vp,f)\, \eta \partial_-{\eta} + 2 \eta' \partial_-{\eta}'\nonumber \\
&  - \frac12 \eta \eta' \eta'' \partial_- \eta \bigg)  + b_0(f) ( f' \partial_- f - f'{}^2 \eta \partial_- \eta ) \Bigg]\,.
\end{align}
In order to compare the $\cN =1$ Chern-Simons action \eqref{S_reducedn=2} with the above action, we can write \eqref{S_reducedn=2}, up to total derivatives, as
\begin{align}
S^{\cN=1} =  \frac{k}{8\pi} \int & \extd t \,\extd \vp \; \bigg[ \Big(\partial_- \log(Y'-k_0 Y-\psi^-\psi^-{}') \Big)\Big( \partial_\varphi \log(Y'-k_0 Y-\psi^-\psi^-{}')\Big)- k_0^2\nn\\
&+4\frac{\Big(\psi^-{}'-\frac{k_0}{2}\psi^-\Big)\Big(\partial_- (\psi^-{}'-\frac{k_0}{2}\psi^-)-\partial_- \log(Y'-k_0 Y-\psi^-\psi^-{}')^{\frac{1}{2}}\Big)}{Y'-k_0 Y - \psi^- \psi^-{}'}\bigg]\,,
\end{align}
where we have considered a single pair of fermionic generators and no $R$-symmetry. We have also set $\eta^{\alpha\beta} = - 2, \umat = 1$ and since there is no $R$-symmetry we have $\lambda^a = 0$.

Like in the bosonic case, both fields $Y(\vp)$ and $\psi^-(\vp)$ are periodic. In order to relate these fields to those parameterizing the super circle, i.e. fields with the periodicity condition of elements of the $\widehat{\rm Diff}(S^{1|1})$, we consider the following field transformations
\begin{align}
   	Y(f(\vp)) &= e^{-k_0 (f(t,\vp)-\vp)},\\ 
   	\psi^-(f(\vp),\eta(\vp)) &= \sqrt{Y'-k_0 Y}\eta(\vp)=\sqrt{-k_0 f'} e^{-\frac{k_0}{2} (f(t,\vp)-\vp)}\eta(\vp),
\end{align}
which can be obtained from 
\begin{align}
    	\tilde{\tilde{\vp}} = e^{-k_0(\tilde{\vp}-\vp)},\qquad\qquad
    	\tilde{\tilde{\theta}} =\sqrt{-k_0 e^{-k_0 (\tilde{\vp}-\vp)}}\tilde{\theta},
\end{align}
with $\tilde{\tilde{\vp}}=Y(\vp+\theta \chi)$.

For the new fields $f(\vp)$ and $\eta(\vp)$ we have
\begin{equation}
    f(\vp+2\pi)=f(\vp)+2\pi,\qquad\qquad \eta(\vp+2\pi) =\eta(\vp),
\end{equation}
while $Y(\vp)$ and $\psi^-(\vp)$ are periodic.

In the new variables, the action takes the form \eqref{N1supervirgeom} with the central charge and the orbit representative given as
\begin{equation}\label{eq:orbit_rep_ident_k0_N1}
    c= 6k,\qquad\qquad\qquad b_0 = \frac{k}{8\pi} k_0^2.
\end{equation}
And so we find the bulk holonomy is still related to the orbit representative $b_0$ as in \eqref{b0L0}.

This proves our earlier claim that $\cN =1$ boundary action on the outer boundary \eqref{S_reduced} obtained from the Hamiltonian reduction in the presence of non-trivial holonomy parametrized with $k_0$ is the geometric action on the coadjoint orbit of group of reparametrization of $S^{1|1}$ where the orbit representative $b_0$ is identified with the holonomy as in \eqref{eq:orbit_rep_ident_k0_N1}.

\subsection{$\cN=2$ supersymmetry}\label{sec:N=2}

Here we will discuss the case of $\cN=2$ supersymmetry. 
The $\cN=2$ supercircle can be studied in a complex basis where
\begin{equation}
\theta = \frac{1}{\sqrt{2}} (\theta^1 + i \theta^2)\,, \qquad \bar{\theta} = \frac{1}{\sqrt{2}} (\theta^1 - i \theta^2)\,,
\end{equation}
Defining the supercovariant derivatives as
\begin{equation}
D = \partial_\theta + \bar{\theta} \partial_\vp \,, \qquad \bar{D} = \partial_{\bar{\theta}} + \theta \partial_\vp\,,
\end{equation}
we have $D^2 = 0 = \bar{D}^2$ and $\{D,\bar{D}\} = 2\partial_\vp$. Equation \eqref{Dtrafo} now implies that
\begin{equation}
D = D\tilde{\theta} \tilde{D}\,, \qquad \bar{D} = \bar{D} \tilde{\bar{\theta}} \tilde{\bar{D}}\,, \quad \text{and} \quad D\tilde{\bar{\theta}} = 0 = \bar{D}\tilde{\theta}
\end{equation}
and $\det{D\tilde{\theta}} = D\tilde{\theta} \bar{D}\tilde{\bar{\theta}}$.

The $\cN=2$ Schwarzian derivative in this notation becomes
\begin{equation}
S^{(2)}(z; \tilde{z}) = \frac{1}{24\pi} \left( \frac{\partial_\vp \bar{D}\tilde{\bar{\theta}}}{\bar{D}\tilde{\bar{\theta}}} - \frac{\partial_\vp D\tilde{\theta}}{D\tilde{\theta}} - 2 \frac{\partial_\vp \tilde{\theta} \partial_\vp \tilde{\bar{\theta}}}{ D\tilde{\theta} \bar{D}\tilde{\bar{\theta}}}\right)
\end{equation}
and the Maurer-Cartan form is $Y =  -\frac{ \extd\tilde{\vp} + \tilde{\theta} \extd \tilde{\bar{\theta}} + \tilde{\bar{\theta}} \extd \tilde{\theta} }{ D\tilde{\theta} \bar{D}\tilde{\bar{\theta}}} \equiv - \frac{ \delta \tilde{l}}{ D\tilde{\theta} \bar{D}\tilde{\bar{\theta}}} $. Useful relations to write \eqref{Omega} as a total exterior derivative are $\extd(\delta \tilde{l}) = 2 \extd \tilde{\theta} \extd \tilde{\bar{\theta}}$, $D(\delta\tilde{l}) = 2 D\tilde{\theta} \extd \tilde{\bar{\theta}}$ and $\bar{D}(\delta \tilde{l}) = 2 \bar{D} \tilde{\bar{\theta}} \extd \tilde{\theta}$. 
The answer for the geometric action of the $\cN=2$ theory is
\begin{align}\label{N2supergeom}
S_{\widehat{\rm Diff}(S^{1|2})} (\tilde{z}; B_0, c) =  \int \extd t \extd z  &\bigg( \frac{c}{24\pi} \left( \frac{\partial_\vp \tilde{\theta} \partial_- \tilde{\bar{\theta}} - \partial_\vp \tilde{\bar{\theta}} \partial_- \tilde{\theta}}{D \tilde{\theta} \bar{D}\tilde{\bar{\theta}}}\right)\nn\\
&+  B_0 ( \tilde{z}) ( \partial_-\tilde{\vp} + \tilde{\theta} \partial_- \tilde{\bar{\theta}} + \tilde{\bar{\theta}} \partial_-\tilde{\theta}) \bigg)\,.
\end{align}
%The Hamiltonian from \eqref{HamgenN} simply becomes
%\begin{equation}
%H = - \int \extd t\extd z \; \left(D\tilde{\theta} \bar{D}\tilde{\bar{\theta}}  B_0 ( \tilde{x}, \tilde{\theta}, \tilde{\bar{\theta}}) - \frac{c}{3\pi} \left( \frac{\partial_x \bar{D}\tilde{\bar{\theta}}}{\bar{D}\tilde{\bar{\theta}}} - \frac{\partial_x D\tilde{\theta}}{D\tilde{\theta}} - 2 \frac{\partial_x \tilde{\theta} \partial_x \tilde{\bar{\theta}}}{ D\tilde{\theta} \bar{D}\tilde{\bar{\theta}}}\right) \right)
%\end{equation}
In order to relate this to the component action coming from the reduction of the Chern-Simons theory we need a suitable parameterization of $\tilde \vp, \tilde \theta$ and $\tilde{\bar \theta}$. We can obtain this by writing the most general diffeomorphism of the $\cN =2$ super-circle compatible with the constraints \eqref{Dtrafo}, which in the complex basis read:
\begin{align}
D \tilde{\bar{\theta}} = 0 \,, \qquad D \tilde \vp = \tilde{\bar{\theta}} D\tilde \theta \,, \\
\bar{D} \tilde \theta = 0 \,, \qquad \bar{D} \tilde \vp = \tilde \theta \bar{D} \tilde{\bar{\theta}} \,.
\end{align}
A general super-reparameterization satisfying these constraints can be written in terms of two bosonic fields: $f(t,\vp)$, which parameterizes the $\widehat{ \rm Diff}(S^1)$ element and satisfies $f(\vp+2\pi) = f(\vp) + 2\pi$ and $ \sigma(t,\vp)$ which is a $U(1)$ $R$-symmetry field. There are also two Grassmann valued fields $\psi_\alpha(t,\vp)$ with $\alpha = {1,2}$,
\begin{subequations}
	\label{N2parameters}
	\begin{align}
	\tilde \vp & = f(t,\vp) -  \theta \psi_1(t,\vp) \bar{g}(t,\vp) + \bar{\theta} \psi_2(t, \vp)g(t,\vp) +  h(t,\vp) \theta \bar{\theta} \,, \\
	\tilde \theta & =  \psi_2(t,\vp) +  \theta \left( \bar{g}(t,\vp) +  \bar{\theta} \psi_2'(t,\vp) \right)\,, \\
	\tilde{\bar{\theta}} & = - \psi_1(t,\vp) +  \bar{\theta} \left( g(t,\vp) - \theta  \psi_1'(t,\vp) \right)\,, 
	\end{align}
\end{subequations}
where
\begin{align}
g(t,\vp) & =  \sqrt{f'(t,\vp)} e^{ i \sigma(t,\vp)/2} \left( 1 - \frac14 \frac{\psi \cdot \psi'}{f'} - \frac{1}{32} \left(\frac{\psi \cdot \psi'}{f'} \right)^2 \right) \\
\bar{g}(t,\vp) & = \sqrt{f'(t,\vp)} e^{- i \sigma(t,\vp)/2} \left( 1 - \frac14 \frac{\psi \cdot \psi'}{f'} - \frac{1}{32} \left(\frac{\psi \cdot \psi'}{f'} \right)^2 \right) \\
h(t,\vp) & = - \frac12 \left( \psi \lambda \psi \right)'
\end{align}
where we remind the reader that $\cdot$ is contraction with $\eta^{\alpha\beta}$ and $ \psi \lambda \psi = \psi_{\alpha} \lambda^{\alpha \beta} \psi_\beta$. Our conventions for $\eta^{\alpha\beta}$ and $\lambda^{\alpha\beta}$ are listed in the appendix \ref{app:conv}.

In order to show that the action obtained from the Hamiltonian reduction is indeed the geometric action, we proceed with a bit different approach.

When the geometric action for $\cN =2$ \eqref{N2supergeom} is expanded using the above parameterization, the $c$-dependent term becomes equal to the action obtained from Chern-Simons theory \eqref{S_reducedn=2} for $\cN=2$ on the outer boundary, when the holonomy contributions set to zero with $\tilde{Y}(t,\vp),\tilde{\psi}^-_\alpha(t,\vp)$ and $\tilde{C}^a(t,\vp)$ replaced by $f(t,\vp), \psi_\alpha$ and $\sigma(t,\vp)$ respectively. That confirms the equivalence of the theories. However, the fields $\tilde{Y}(t,\vp)$, $\tilde{\psi}^-_\alpha(t,\vp)$ and $\tilde{C}^a(t,\vp)$ are no longer periodic. The amount of non-periodicity is encoded in the phase that these fields pick up under the action of the holonomy on the group elements.  All there is to do now is to find the transformation which reinstates the $B_0(\tilde{z})$ term and read of the relation between the zero modes and the orbit representative $B_0$. To this end we should solve \eqref{orbitredef} for $\cN =2$. The orbit representative has only non-zero bosonic values, which we will take to be constant.
\begin{equation}
B_0(\tilde{z}) = b_\sigma + \tilde{\theta} \tilde{\bar{\theta}} b_0 \,.
\end{equation}
A super-reparameterization which solves \eqref{orbitredef} is
\begin{equation}\label{N2superredef}
\tilde{\tilde{\vp}} = e^{-\mu \tilde{\vp}} \,, \qquad \tilde{\tilde{\theta}} = e^{\frac{12\pi b_{\sigma}}{c} \tilde{\vp}} \tilde{\theta} \sqrt{- \mu e^{-\mu \tilde{\vp}}} \,, \qquad \tilde{\tilde{\bar{\theta}}} = e^{-\frac{12\pi b_{\sigma}}{c} \tilde{\vp}} \tilde{\bar{\theta}} \sqrt{- \mu e^{-\mu \tilde{\vp}}} \,, 
\end{equation}
where now
\begin{equation}
\mu = \sqrt{48\pi \left( \frac{b_0}{c} + 12\pi \frac{b_{\sigma}^2}{c^2} \right) }
\end{equation}
This transformation relates the different components of $\tilde{\tilde{z}}$ to the components of $\tilde{z}$ as given in \eqref{N2parameters}. If one parameterizes $\tilde{\tilde{z}}$ in the same way, but with fields $\tilde{Y}(t,\vp), \tilde{C}(t,\vp)$ and $\tilde{\psi}^-(t,\vp)$ instead of $f(t,\vp), \sigma(t,\vp)$ and $\psi_\alpha$, then the field redefinitions which reinstate the orbit representative term, as deduced from \eqref{N2superredef} are
\begin{subequations} \label{N2componentredef}
	\begin{align}
	\tilde{Y}(t,\vp) & = e^{- \mu f(t,\vp)} \,, \\
	\tilde{C}(t,\vp) & = \sigma(t,\vp) + \frac{24\pi i b_\sigma}{c} f(t,\vp) + \mu \psi_1(t,\vp) \psi_2(t,\vp) \,,\\
	\tilde{\psi}^-_1(t,\vp) &= e^{-\frac{12\pi b_\sigma}{c} f(t,\vp)}\sqrt{-\mu e^{-\mu f(t,\vp)}} \psi_1(t,\vp) \\
	\tilde{\psi}^-_2(t,\vp) &= e^{\frac{12\pi b_\sigma}{c} f(t,\vp)}\sqrt{-\mu e^{-\mu f(t,\vp)}} \psi_2(t,\vp)
	\end{align}
\end{subequations}
The periodicities of the fields $\tilde{Y}(t,\vp), \tilde{C}(t,\vp)$ and $\tilde{\psi}^-(t,\vp)$ as dictated by the holonomy of the Chern-Simons connection are for $\cN =2$
\begin{subequations}
	\label{periodicityN2}
\begin{align}
\tilde{Y}(\vp + 2\pi) &= e^{-2\pi \sqrt{4\cL_0}} \tilde{Y}(\vp)\,,  \\
\tilde{C}(\vp + 2\pi) & = \tilde{C}(\vp) - 2\pi i \cB_0 \\
\tilde{\psi}^-_\alpha(\vp + 2\pi) &= e^{-\pi \sqrt{4\cL_0}} \tilde{\psi}^-_\beta(\vp) e^{-2\pi \cB_0 \lambda^\beta{}_\alpha}\, ,
\end{align}
\end{subequations}
This is compatible with \eqref{N2componentredef} whenever
\begin{equation}\label{N2orbitrep}
\hat{\cL_0} = \cL_0 - \frac14 \cB_0^2 = \frac{12 \pi}{c} b_0 \,, \qquad \cB_0 = - \frac{24\pi}{c} b_\sigma
\end{equation}
So we see that the orbit representative $b_0$ corresponds to the zero-mode of the Sugawara-shifted stress tensor $\hat{\cL}$ \eqref{chargeL}.

One can now go back to the original periodic fields $Y(t,\vp), C(t,\vp)$ and $\psi^-(t,\vp)$ with the following field redefinitions
\begin{subequations}\label{eq:periodicfieldsN2}
\begin{align}
Y(\vp) &= e^{\mu \vp} \tilde{Y}(t,\vp)\,,  \\
C(\vp) & = \tilde{C}(t,\vp) +i\cB_0 \vp, \\
\psi^-_\alpha(\vp) &= \tilde{\psi}^-_\beta(t,\vp)  e^{\frac{\mu}{2}\vp} e^{\cB_0 \lambda^\beta{}_\alpha \vp}.
\end{align}
\end{subequations}
Here one sets
\begin{align}
    \mu = \sqrt{k_0^2+k_r^2},
\end{align}
where 
\begin{equation}
    b_0 = \frac{k}{8\pi}k_0^2,\qquad\qquad b_{\sigma}^2 = \frac{k^2}{16\pi^2} k_r^2.
\end{equation}
We have now shown explicitly that the Hamiltonian reduction of Chern-Simons theory for the $Osp(2|2)$ group under the highest-weight boundary conditions \eqref{conn-hwg} gives a two dimensional field theory equivalent to the geometric action of the centrally extended group of diffeomorphisms of the $\cN =2$ supercircle on its coadjoint orbit. The orbit representatives are related to the zero modes of the Chern-Simons charges by \eqref{N2orbitrep}.

\section{Conclusions}

The main result of this paper is to address in a precise and detailed way the boundary actions for extended supergravity and higher spin $W_3$ gravity in three dimensions in the presence of non-trivial bulk holonomies. We provide the complete analysis where the topology of the manifold is $\mathbb{R}\times S^1 \times [0,1]$. The spatial surfaces with the annulus topology $S^1 \times [0,1]$ are then describing an AdS$_3$ space with two boundaries. Even though we discuss the details for this case, the extension to the topologies with more boundaries is straightforward.

Similar to three dimensional Chern-Simons pure gravity, the boundary actions both for extended supergravity and higher spin $W_3$ gravity are written in terms of two free chiral theories living on each boundary, coupled through the zero modes of bulk holonomy which constitutes global degrees of freedom. These so-called Wilson lines are stretched between the boundaries and can be considered as the wormholes in the bulk. This is indeed the reason that the Hilbert space of the quantized boundary field theory is not just a cross product of the states on two boundaries, the so called factorization problem in gravity. 

The novelty of this work is that we are able to write down the boundary action, including the boundary Hamiltonian, as a generalization of the Schwarzian action in the cases where the asymptotic symmetry algebra is non-linear. In this situation, the prescription of writing the geometric action based on the coadjoint representation is no more accessible. The action on the boundary must be seen as the action on the symplectic leaves of a Poisson manifold of the corresponding Poisson algebra of the phase space variables. The Hamiltonian reduction in each case will result in the action on the symplectic leaves. In the case that the asymptotic symmetry algebra is linear, the symplectic leaves are coadjoint orbits and the result coincides with the geometric action on the coadjoint orbits. Moreover, we discussed the invariance of the boundary actions in each case. The action is invariant under $\rm{Diff}(S^{1|1})/S^1$ and $\rm{Diff}(S^{2|2})/S^1\times S^1$ for $\cN =1$ and $\cN=2$ AdS$_3$ supergravity respectively. The boundary actions of the $W_3$ gravity were also shown to be invariant under $\rm{Diff}(\rm{RP}^2)/S^1\times S^1$. 

The boundary action of $W_3$ theory in the diagonal $\mathfrak{sl}(2)$ embedding has striking similarities with $\cN =2$ supergravity Schwarzian action. The difference, that can also be inferred from the difference between $\mathfrak{sl}(3)$ algebra in this representation and $\cN =2$ Neveu-Schwarz supergravity algebra, is that in the case of $\mathfrak{sl}(3)$ we have two fermionic fields with bosonic statistics, while in the supergravity the fermions anticommute. This resemblance gets more interesting when one realizes the isomorphism $\rm{RP}^{2|2}\cong S^{2|2}$ with $S^{2|2}$ being the double covering space. Finding an appropriate way to go from $\rm{RP}^{2|2}$ to $\rm{RP}^{2}$ to make a link between these two theories is an interesting question given that one is able to construct the geometric action via the Kirilov-Kostant construction for $\cN =2$ supergravity.

Recent studies pointed to the crucial role of the Schwarzian action in the computation of the partition functions of 2D and 3D gravity, see \cite{Mertens:2018fds} and references therein. It would be interesting to compute partition functions for the Schwarzian actions obtained in this work. The advantage here is that not only these actions control the solution space via the (constant) holonomies, but also they contain information on global degrees of freedoms related to the bulk holonomy. These holonomies must be then considered as dynamical degrees of freedom that have to be varied in the action. They also appear in the measure of the partition function and must be integrated over. This matches with the expectation that in obtaining the partition function of a quantum theory of gravity by performing a path integral on a classical theory one expects to sum over all the solutions in the solution space of the theory. This line of thought will be pursued elsewhere.

\section*{Acknowledgements}
The authors would like to extend their gratitude to Marc Henneaux who has played an important role through collaboration and discussion at all the stages of this work. TN would like to thanks Universit\'e Libre de Bruxelles and International Solvay Institutes for support during the initial stages of the work. AR would like to thank the department of theoretical and mathematical physics of Universit\'e Libre de Bruxelles for their hospitality in various stages of this work. The work of TN is supported by ANID FONDECYT grant No. 3220427. The work of AR is supported by the Croatian Science Foundation project IP-2019-04-4168. 

\appendix

%%%%%%%%%%%%%%%%%%%%%%%%%%%%%%%
%%%%%%%%%%%%%%%%%%%%%%%%%%%%%%%
\section{Conventions}\label{app:conv}
In this appendix we list some conventions used throughout the text

\subsection{Matrix Representation for $Osp(2|2)$}
For the sake of simplicity and in order to be able to have an explicit calculation we restrict ourselves to $\mathcal{N}=2$ supersymmetry. In that case, the even subalgebra $\mathfrak{g}_0= \mathfrak{sl}(2,\mathbb{R}) \oplus \mathfrak{so}(2)$. The superalgebra $\mathfrak{osp}(2|2,\mathbb{R})$ is generated with 8 generators of which 4 are generators of $\mathfrak{g}_0$. We can write explicitly\footnote{A generic form of matrix representation of $\mathfrak{osp}(2\ell+1|2,\mathbb{R})$  generators can be written as
	\bea
	\begin{pmatrix} 
		\begin{matrix}
			a & b & u \\
			c & -a^T & v \\ -v^T & -u^T & 0 \end{matrix} & \rvline & \begin{matrix}
			x & x_1 \\ y & y_1 \\ z & z_1 \end{matrix} \\ 
		\hline 
		\begin{matrix}
			y_1^T & x_1^T & z_1^T \\ -y^T & -x^T & -z^T \end{matrix} & \rvline & SL(2) \end{pmatrix},
	\eea
	where $a$ is any $\ell \times \ell$ matrix, $b,c$ are antisymmetric $\ell \times \ell$ matrices, $u,v$ are $\ell \times 1$ column matrices, $x,x_1,y,y_1$ are $\ell \times 1$ column matrices and $z,z_1$ are real numbers.
	A generic form of matrix representation of $\mathfrak{osp}(2\ell|2,\mathbb{R})$  generators can be obtained by eliminating the middle row and column of the $\mathfrak{osp}(2\ell+1|2,\mathbb{R})$ matrix forms, i.e.
	\bea
	\begin{pmatrix} 
		\begin{matrix}
			a & b \\
			c & -a^T  \end{matrix} & \rvline & \begin{matrix}
			x & x_1 \\ y & y_1 \end{matrix} \\ 
		\hline 
		\begin{matrix}
			y_1^T & x_1^T  \\ -y^T & -x^T  \end{matrix} & \rvline & SL(2) \end{pmatrix}.
	\eea
}

\bea
L_0&=& \frac{1}{2} \begin{pmatrix} \bigzero & \rvline & \bigzero \\ \hline \bigzero & \rvline & \begin{matrix}
		1 & 0 \\ 0 & -1 \end{matrix}\end{pmatrix},\qquad
L_{+}= \begin{pmatrix} \bigzero & \rvline & \bigzero \\ \hline \bigzero & \rvline & \begin{matrix}
		0 & 1 \\ 0 & 0 \end{matrix}\end{pmatrix},\qquad
L_{-}= \begin{pmatrix} \bigzero & \rvline & \bigzero \\ \hline \bigzero & \rvline & \begin{matrix}
		0 & 0 \\ 1 & 0 \end{matrix}\end{pmatrix},\nonumber\\
Q_{+}^1 &=& \begin{pmatrix} \bigzero & \rvline & \begin{matrix}
		0 & -1 \\ 0 & 0 \end{matrix} \\ \hline \begin{matrix}
		0 & -1 \\ 0 & 0 \end{matrix} & \rvline & \bigzero \end{pmatrix},\qquad
Q_{-}^1 = \begin{pmatrix} \bigzero & \rvline & \begin{matrix}
		1 & 0 \\ 0 & 0 \end{matrix} \\ \hline \begin{matrix}
		0 & 0 \\ 0 & -1 \end{matrix} & \rvline & \bigzero \end{pmatrix},\nonumber\\
Q_{+}^2 &=& \begin{pmatrix} \bigzero & \rvline & \begin{matrix}
		0 & 0 \\ 0 & -1 \end{matrix} \\ \hline \begin{matrix}
		-1 & 0 \\ 0 & 0 \end{matrix} & \rvline & \bigzero \end{pmatrix},\qquad
Q_{-}^2 = \begin{pmatrix} \bigzero & \rvline & \begin{matrix}
		0 & 0 \\ 1 & 0 \end{matrix} \\ \hline \begin{matrix}
		0 & 0 \\ -1 & 0 \end{matrix} & \rvline & \bigzero \end{pmatrix},\nonumber\\
T &=& \frac{1}{2}\begin{pmatrix} \begin{matrix}
		1 & 0 \\ 0 & -1 \end{matrix} & \rvline & \bigzero \\ \hline \bigzero & \rvline & \bigzero \end{pmatrix}.
\eea
These generators satisfy the algebra \eqref{bracket-init}-\eqref{ospN} with $\alpha=1,2$, $d=2$, $C_{\rho}=-1/4$. Also we have
\begin{align}
\eta^{\alpha\beta}&=\begin{pmatrix}
0 & 2 \\ 2 & 0
\end{pmatrix},\qquad \quad \eta_{\alpha\beta}=\begin{pmatrix}
0 & \frac{1}{2} \\ \frac{1}{2} & 0
\end{pmatrix},\\
\lambda^{\alpha\beta}&=\begin{pmatrix}
0 & -1 \\ 1 & 0
\end{pmatrix},\qquad \lambda^\alpha_\beta =\begin{pmatrix}
-\frac{1}{2} & 0 \\ 0 & \frac{1}{2}
\end{pmatrix}.
\end{align}

%%%%%%%%%%%%%%%%%%%%%%%%%%%%%%%%%%%%%%%%%%%%%%%%%%%%%%%%%%%%%%%%%%%%%

%%%%%%%%%%%%%%%%%%%%%%%%%%%%%%%%%%%%%%%%%%%%%%%%%%%%%%%%%%%%%%%%%%%%%
\section{Action on the symplectic leaves}\label{app:sym_leaves}

The Poisson algebra of charges of $W_3$ gravity, which was discussed in section \ref{sec:Higher-Spin}, upon reduction forms a finite dimensional algebra known as finite $W$-algebra. In the case of the principal embedding, we will end up with an Abelian algebra. In the case of the diagonal embedding, it forms a 4-dimensional algebra known as $w_3^{(2)}$, however, it has a trivial one dimensional unitary representation. The symplectic leaves in this case are determined by the intersection of coadjoint orbits of $\mathfrak{sl}(3,\mathbb{R})$ and $\mathcal{G}_{lw}=\left\{x\in \mathfrak{sl}(3)\,|\,[U_{+1},x]=0\right\}$. It was shown in \cite{deBoer:1995cqx} that it is topologically equivalent to a two dimensional surface in $\mathbb{C}^3$ given by the equation
\begin{equation}
    z_2z_3 = c_2 -2 c_1 z_1 +z_1^3,
\end{equation}
where $c_1$, $c_2$ are two arbitrary constants.

The symplectic form on the symplectic leaves is not obtained trivially in the case of non-linear algebras. This has to do with the fact that the Cartan formula which is used to define the Schwarzian in the linear algebras is no longer providing an expression for the Schwarzian \cite{Michel:2007}. This is the case for superalgebras with $\mathcal{N}>2$ and $W_N$ algebras with $N>2$.

There exists a tower of Poisson algebras (Gel'fand-Dikii brackets) on linear differential operators of an arbitrary order with periodic coefficients on the circle. The first algebra in this hierarchy is the Virasoro algebra. It is known that in that case the question of classification of symplectic leaves of a Poisson manifold with the Gel'fand-Dikii algebra, is equivalent to the classification i) of orbits of the coadjoint representation of the Virasoro group, ii) of normal forms of Hill equations, and iii) of types of projective structures on the circle \cite{10.1307/mmj/1028989908,Lazutkin1975,10.1007/BFb0066026,Segal1981}. 

Similarly and along the same line, the classification of contact-projective structures on a supermanifold $\rm{RP}^{2|\cN}$ and its
double covering $S^{2|\cN}$ was discussed in \cite{Ovsienko_1989}. For $\mathcal{N}<3$ it is equivalent to the classification of the orbits of the coadjoint representation of a Lie superalgebra of NS-R type and the supersymmetric analogue of the Hill equation.

What all these cases have in common is that the classification of symplectic leaves is connected  with the computation of homotopy classes of nondegenerate curves on $S^1$, $\rm{RP}^{2|\cN}$ or $S^{2|\cN}$. The monodromy operator is the only local invariant of a symplectic leaf of
the Gel'fand-Dikii bracket associated with one of the classical groups and it determines the conjugacy class in the corresponding matrix Lie group.

In the case of $\mathcal{N}>2$, the differential equation will turn to a pseudo-differential equation. For $\mathcal{N}=3$, a solution to such an equation was found \cite{RADUL1990} and therefore one could write the Schwarzian, however it is not possible in any general case to find a solution. 

All these make it difficult (if not impossible) to write down the action in the case of a non-linear algebras. We claim that the boundary actions derived in Sections \ref{sec:reduction} and \ref{sec:Higher-Spin}, obtained through Hamiltonian reduction, should be considered as the actions on the symplectic leaves which otherwise are difficult to obtain. We provide a practical approach to find these actions while the geometric approach seems to be much more involved.

\section{Geometric quantization for (super)conformal groups}\label{app:geomquan}

In this section, we provide a short review of geometric quantization for (super)conformal group based on \cite{Alekseev:1988ce,Alekseev:1990mp,Aratyn:1989qq,Aratyn:1990dj,Barnich:2017jgw}.

\subsection{Geometric actions on the coadjoint orbit}\label{sec:geomaction}
For any Lie group $G$ with Lie algebra $\mathfrak{g}$, the adjoint action of $G$ on $\mathfrak{g}$ is
\begin{equation}
	\Ad_g X = g X g^{-1}\,.
\end{equation}
The coadjoint action of $G$ on the dual space $\mathfrak{g}^*$ is defined as
\begin{equation}
\langle \Ad_{g^{-1}}^*  b, X \rangle = \langle b,\Ad_{g} X \rangle\,,
\end{equation}
where $b \in \mathfrak{g}^*$ and $\langle\, , \, \rangle$ is the pairing between $\mathfrak{g}$ and $\mathfrak{g}^*$. 

For a fixed element $b_0$ of $\mathfrak{g}$ the coadjoint action of $G$ spans the orbit $\cO_{b_0}$, defined as the set of elements $b \in \mathfrak{g}^*$ such that
\begin{equation}
b = \Ad_{g^{-1}}^* \, b_0\,.
\end{equation}
Coadjoint orbits are symplectic manifolds isomorphic to $G/H_{b_0}$ where $H_{b_0}$ is the stabilizer subgroup of the orbit, i.e. all elements $h\in G$ such that 
$\Ad^*_h \, b_0 = b_0$.
The symplectic form on the coadjoint orbit is the Kirillov-Kostant symplectic form and it is defined by
\begin{equation}\label{KKform}
	\Omega = \frac12 \langle b, \ad_{Y} Y \rangle\,.
\end{equation}
here $\ad_Y$ is the adjoint action of $\mathfrak{g}$ on itself and $Y$ is obtained as the solution to
\begin{equation}\label{Yeqn}
\extd b = - \ad_Y^* b
\end{equation}
For $b = \Ad_{g^{-1}}^* b_0$ this gives
\begin{equation}\label{Yeqn2}
\extd (\Ad_{g^{-1}}^* b_0)  = - \ad_Y^* (\Ad_{g^{-1}}^* b_0)
\end{equation}
From \eqref{Yeqn} it follows that $Y$ solves the Maurer-Cartan equation $\extd Y = - \frac12 \ad_Y Y$, whose solution is locally $Y = g^{-1}dg$. In practice it will also be useful to obtain $Y$ from \eqref{Yeqn2}.

An action whose phase space coincides with the coadjoint orbit can be obtained by writing the Kirillov-Kostant symplectic form \eqref{KKform} as a total exterior derivative
\begin{equation}
\Omega = d\alpha\,.
\end{equation}
The geometric action on the coadjoint orbit is then obtained by integrating the presymplectic potential $\alpha$ over the orbit
\begin{equation}
I [g;b_0] = \int_\gamma \alpha = \int_{\gamma} \langle \Ad_{g^{-1}}^* b_0, Y \rangle
\end{equation}
where $\gamma$ parameterizes a curve on the orbit $\cO_{b_0}$.

We are interested in the geometric action for centrally extended Lie groups, such as the superconformal groups. A centrally extended group $\hat{G} =G \times \bR$ with Lie algebra $\hat{\mathfrak{g}}$ has elements $(X,n) \in \hat{\mathfrak{g}}$ and $(b,c) \in \hat{\mathfrak{g}}^*$ such that the bilinear form reads
\begin{equation}
\langle (b,c) , (X,n) \rangle = \langle b, X \rangle + c n
\end{equation}
The adjoint and coadjoint action of $\hat{G}$ now reads:
\begin{align}
\Ad_g (X,n) & = (\Ad_g X, n - \langle S(g),X \rangle ) \,, \\
\Ad_g^* (b,c) & = (\Ad_g^* b - c S(g^{-1}), c)\,.
\end{align}
Here $S(g)$ is the Souriau cocycle, satisfying the condition
\begin{equation}\label{cocycle}
	S(g_1 g_2) = \Ad_{g_2^{-1}}^* S(g_1) + S(g_2) \,,
\end{equation}
together with $S(I) = 0$, such that $S(g) = - \Ad_{g^{-1}}^* S(g^{-1})$. The adjoint action of $\mathfrak{g}$ on itself is
\begin{equation}
\ad_{(X_1,n_1)}(X_2,n_2) =  [(X_1,n_1),(X_2,n_2)] = ([X_1,X_2], - \langle s(X_1) ,X_2 \rangle )
\end{equation}
where $s(X)$ is the infinitesimal limit of $S(g)$.

The coadjoint action on $\mathfrak{g}$ is defined correspondingly as $\langle \ad_{(X_1,n_1)}^*(b,c),(X_2,n_2)\rangle = - \langle (b,c), \ad_{(X_1,n_1)}(X_2,n_2)\rangle$ and reads
\begin{equation}
\ad_{(X,n)}^* (b,c) = (\ad^*_X b + c\, s(X), 0)\,.
\end{equation}

The Kirillov-Kostant symplectic form now becomes
\begin{equation}\label{KKform2}
\Omega = \frac12 \langle (b,c), [(Y,n_Y), (Y,n_Y)] \rangle =  d  \langle \Ad^*_{g^{-1}} (b_0,c), (Y, n_Y) \rangle
\end{equation}
where $n_Y$ solves
\begin{equation}\label{nYeqn}
\extd n_Y =  \frac12 \langle s(Y), Y \rangle\,.
\end{equation}
This makes the geometric action on the coadjoint orbit of a centrally extended group
\begin{equation}
I[g;b_0,c] = \int_{\gamma} \alpha  = - \int_{\gamma} \langle \Ad^*_{g^{-1}} (b_0,c), (Y, n_Y) \rangle\,.
\end{equation}
An alternative way to obtain the geometric action on the coadjoint orbit is to use the identity
\begin{equation}\label{dSid}
\extd S(g) = - \ad_Y^* S(g) + s(Y)
\end{equation}
and write the Kirillov-Kostant symplectic form \eqref{KKform2} as:
\begin{equation}\label{KKform3}
\Omega = \extd \alpha  =  d  \langle \Ad^*_{g^{-1}} b_0 , Y \rangle - \frac{c}{2} \langle \extd S(g), Y \rangle \,.
\end{equation}
So to find the geometric action we can either find $(Y,n_Y)$ by solving \eqref{Yeqn2} and \eqref{nYeqn}, or alternatively, we may solve \eqref{Yeqn2} for $Y$ and write the last term in \eqref{KKform3} as a total exterior derivative. 

\subsection{Hamiltonians}

The geometric action on the coadjoint orbit only gives the symplectic part of the action. The evolution on the orbit is determined by adding a suitable Hamiltonian. In \cite{Barnich:2017jgw} it was shown how to add Hamiltonians in such a way as to preserve the gauge symmetries (generated by the stabilizer subgroup) on the orbit. One can do so by adding the Noether charge associated to a global symmetry as the Hamiltonian. 

Suppose the geometric action has a global symmetry generated by a left invariant vector field $V_{(X,n)} = (g X, n)$, satisfying $\mathfrak{L}_{V_{(X,n)}} \alpha = 0$, where $\mathfrak{L}_{V_{(X,n)}} $ is the Lie derivative. Then
\begin{equation}
i_{V_{(X,n)}} \Omega = d Q_{(X,n)}\,, \qquad \text{with: } \;\; Q_{(X,n)} = - \langle (b,c), (X,n) \rangle \,.
\end{equation}
Here $Q_{(X,n)}$ is the Noether charge associated to the global symmetry generated by $V_{(X,n)}$. The Noether charge can be added to the geometric action a Hamiltonian without changing the gauge symmetries generated by the little group on the orbit $H_{b_0}$. 
\begin{equation}\label{IgeomHam}
I[g;b_0,c,H_{(X,n)}] = I[g;b_0,c] - \int_{\gamma} Q_{(X,n)} \extd t =  \int_{\gamma} \langle \Ad^*_{g^{-1}} (b_0,c), (Y, n_Y) + (X,n) \extd t \rangle\,.
\end{equation}

\providecommand{\href}[2]{#2}\begingroup\raggedright\endgroup

\end{document}